\documentclass{aa} 

\usepackage{graphicx}
\usepackage{natbib}
\usepackage{txfonts}
\usepackage{xcolor}
\usepackage{hyperref}
\usepackage{dblfloatfix}

\begin{document} 
   \title{Invariant manifolds in barred galaxy simulations}

   \subtitle{III. Self-regulated weakening of strong spiral arms}

   \author{T. Soler-Terricabras\inst{1}\fnmsep\inst{2}\fnmsep\inst{3},
          S. Roca-Fàbrega\inst{4}, \and
          M. Romero-Gómez\inst{1}\fnmsep\inst{2}\fnmsep\inst{3}
          }

     \titlerunning{Invariant manifolds in barred galaxy simulations -- III}
    \authorrunning{Soler-Terricabras et al.}

   \institute{Departament de Física Quàntica i Astrofísica (FQA), Universitat de Barcelona (UB), c. Martí i Franquès, 1, 08028 Barcelona, Spain\\
              \email{tsoler@fqa.ub.edu}
        \and
             Institut de Ciències del Cosmos (ICCUB), Universitat de Barcelona (UB), c. Martí i Franquès, 1, 08028 Barcelona, Spain
        \and
             Institut d’Estudis Espacials de Catalunya (IEEC), Edifici RDIT, Campus UPC, 08860 Castelldefels (Barcelona), Spain
        \and 
             Lund Observatory, Division of Astrophysics, Department of Physics, Lund University, SE-221 00 Lund, Sweden
             }

   \date{Received 8 June 2026 / Accepted 30 July 2026}

  \abstract
   {Several theories have attempted to explain the formation, evolution, and overall nature of spiral arms in barred galaxies. However, although some are supported by theoretical arguments or numerical simulations, none has been fully conclusive. One such framework is invariant manifolds theory, which links spiral structure to the orbital dynamics associated with the saddle-unstable Lagrangian points. In particular, it has been shown that the fraction of particles trapped by the exterior unstable manifold branches closely follows the temporal evolution of spiral-arm strength.}
   {We investigated the origin of the weakening of strong spiral arms, seeking the physical mechanism responsible for the decline in the fraction of trapped particles over time.} 
   {We reconstructed the gravitational potential of a fully self-consistent $N$-body simulation of a barred galaxy using \texttt{AGAMA}, computed the Lagrangian points, and quantified how their spiral-induced angular displacement affects the geometry of the invariant manifolds and the associated orbital flows.}
   {We find that {sufficiently strong, self-gravitating spiral arms significantly reshape the gravitational potential of a barred galaxy, displacing the equilibrium points} up to $20-50^\circ$ from the bar major axis, where they are expected to lie in the standard invariant-manifold framework. This shift alters the configuration of the manifold branches and disrupts the coherent stellar flows that sustain the spiral structure. As the spiral weakens, the equilibrium points gradually return to their standard configuration, restoring the conditions for manifold-driven spiral arm formation on a timescale of $\sim 200$ Gyr.}
   {We show {for the first time} that invariant manifolds and the spiral structures they generate are coupled through a self-regulating feedback mechanism, {allowing the spiral pattern to recur over time. Our results therefore indicate that, although barred galaxies may sustain spiral activity over long timescales, individual episodes of strong spiral arms are intrinsically transient.}
}


   \maketitle

   \defcitealias{TST2026a}{Paper~I}
   \defcitealias{TST2026b}{Paper~II}

\section{Introduction}\label{sec:introduction}
The formation of spiral arms in barred galaxies has been extensively studied over the past decades; however, far fewer theoretical frameworks are able to simultaneously account for their origin, nature, and recurrent evolution. 
Numerical simulations of {disc galaxies, both barred and unbarred,} consistently show that spiral arms are not steady features, but rather transient structures that can even repeatedly break apart and reform over time \citep[e.g.][]{SellwoodCarlberg1984, CarlbergFreedman1985, Rautiainen1999, Rautiainen2000, Wada2011, Sellwood2011, Grand2012b, Roskar2012, Baba2013, RocaFabrega2013}. 
In recent years, some studies have {sought to explain this recurrent behaviour by proposing mechanisms from a variety of theoretical perspectives} \citep[e.g.][]{SellwoodCarlberg2014, SellwoodCarlberg2019, SellwoodMasters2022, Meidt2025, Meidt2026}.
{Although these studies invoke different physical mechanisms, they consistently indicate that spiral structure is generally non-steady. 
In particular, even if individual spiral arms remain identifiable over long timescales, their amplitudes can evolve substantially with time.}
Achieving a unified framework that simultaneously captures the origin, nature, and long-term evolution of spiral arms is therefore an open challenge for any comprehensive theory of spiral structure.

Among the proposed frameworks, the theory of invariant manifolds \citep{MRG2006, MRG2007, Voglis2006a} has proven particularly successful in describing the formation of spiral and ring structures in barred galaxies. 
In Papers I and II of this series \citep{TST2026a, TST2026b}, we demonstrated that invariant manifolds provide a robust dynamical framework to account for spiral arms formation in a fully self-consistent $N$-body simulation. 
In particular, we showed that a significant fraction of particles in the spiral arms is dynamically linked to manifold-guided orbits, and that their contribution closely tracks the strength of the spiral pattern. 
These results support a picture in which invariant manifolds constitute the dynamical backbone of spiral structure in barred galaxies.

Despite this progress, some fundamental questions remain unanswered: if invariant manifolds are able to sustain spiral structure, why do strong spiral arms {in numerical simulations systematically} weaken?  
Can they be long-lived?
What physical mechanism {underlies the weakening of strong spiral arms and the recurrence of manifold-guided spiral structures}?
In particular, the standard manifold framework primarily addresses the \textit{generation} of spiral structure, implicitly assuming that the underlying potential is dominated by the galactic bar.
Several studies have extended this framework by considering barred--spiral potentials in which the spiral perturbation is included explicitly in the gravitational potential \citep[e.g.][]{Patsis2006, Patsis2009, Patsis2010, TsigaridiPatsis2013, TsigaridiPatsis2015, PatsisTsigaridi2017, Efthymiopoulos2020, Efthymiopoulos2024}. 
However, {in most of the prescribed potentials explored in these studies, the spiral perturbation remains weaker than the bar}. 
By contrast, observations show that many barred galaxies host spiral arms whose strength is comparable to that of the bar, as found in several nearby barred galaxies, including NGC~289, NGC~986, NGC~1300, NGC~1566, NGC~3359, NGC~4535, NGC~5248, NGC~6140, and NGC~7479 \citep[e.g.][]{Buta2015,Font2019,DiazGarcia2019, Lee2023, Williams2024}. 
Such a regime is particularly relevant because the spiral arms themselves may significantly modify the invariant manifolds that support them.
The dynamical consequences of this mutual interaction between the spiral perturbation and the manifold structure have not yet been explored in fully self-consistent simulations.

The main goal of this paper is therefore to investigate how the mutual coupling between the bar and strong spiral arms reshapes the invariant-manifold framework, and to characterise the transition from a regime in which the spiral perturbation is comparable to the bar to a bar-dominated regime, identifying the mechanisms responsible for this evolution. 
We aim to establish whether invariant manifold theory can account not only for the formation of spiral arms, but also for their subsequent weakening in a fully self-consistent system.\\

This paper is organised as follows. 
In Sec.~\ref{sec:methodology}, we briefly describe the simulation and the methodology used to reconstruct the gravitational potential and locate the equilibrium points.
In Sec.~\ref{sec:distortion}, we quantify the effect of spiral arms on the angular displacement of the Lagrangian points with respect to the bar orientation. 
In Sec.~\ref{sec:weakening}, we analyse the impact of these potential distortions on the spiral structure through an analysis of the invariant manifold branches configuration (Sec.~\ref{sec:branches}) and their impact on the spiral evolution (Sec.~\ref{sec:inward_flows}).
Finally, we discuss the implications of our results in Sec.~\ref{sec:discussion} and present our conclusions in Sec.~\ref{sec:conclusions}.
   
\section{Methodology}\label{sec:methodology}
    
In this work, we adopted the same numerical setup and analysis framework as in \citetalias{TST2026a} and \citetalias{TST2026b}, to which we refer for a detailed description. 
We analysed the B1 simulation of \citet{RocaFabrega2013}, a pure $N$-body model of an isolated barred galaxy consisting of $10^6$ stellar particles and a total disc mass of $M_{d}=5.0\times10^{10}\,M_\odot$.
This simulation has been extensively analysed in a number of previous studies, providing a detailed understanding of the structural properties and time evolution of both the bar and spiral patterns. 
Although its resolution is not state-of-the-art, it is sufficient to robustly trace large-scale non-axisymmetric structures, as demonstrated by convergence tests presented in the original work. Building on this well-characterised framework, our goal is to establish a solid baseline for the present analysis, with the prospect of extending this approach to high-resolution cosmological simulations in future work.

From this simulation, we selected a sequence of snapshots spanning $561\,\mathrm{Myr}$, corresponding to the early evolutionary phase following bar formation and stabilisation, in which both the bar and the spiral structure are well developed.
As shown in Fig.~1 of \citetalias{TST2026a}, the bar remains dynamically dominant throughout this interval, with a nearly constant pattern speed, while the two bisymmetric spiral arms exhibit significant temporal variations in amplitude. 
This makes the dataset particularly suitable for investigating the interplay between spiral structure and the underlying gravitational potential.
The gravitational potential of each snapshot was reconstructed using the same procedure as in \citetalias{TST2026a}, combining the snapshot-characterisation method of \citet{Dehnen2022} with the \texttt{AGAMA} software \citep{Vasiliev2019}. 
The resulting smooth, time-dependent potentials allow us to accurately determine the location of the equilibrium (Lagrangian) points and to compute the associated invariant manifolds following the formalism of \citet{MRG2006, MRG2007}.

In contrast to \citetalias{TST2026a} and \citetalias{TST2026b}, where the focus was on the orbital content and trapping properties of disc particles, here we concentrated on the geometry of the phase-space structures relative to the large-scale features of the galaxy. 
In particular, we measured the angular displacement of the Lagrangian points with respect to the bar axes and analysed how these shifts correlate with the strength of the spiral component. 
We then examined the corresponding changes in the morphology of the invariant manifolds and their impact on the large-scale structure of the disc.
As in the previous papers, the subsequent analysis has been restricted to particles close to the disc midplane ($|z|<400\,\mathrm{pc}$) and with small vertical velocities ($|v_z|<20\,\mathrm{km}\,\mathrm{s}^{-1}$) ensuring that the dynamics is dominated by planar motion.
This approximation provides a robust and appropriate framework for the present analysis: the simulated galaxy is isolated, and previous studies have demonstrated that the main invariant-manifold-driven orbital structures are largely preserved in the planar limit \citep[e.g.][]{OllePfenniger1998, KatsanikasPatsis2022, HarsoulaKatsanikas2025}.
    
\section{Spiral arms distortion of the potential}\label{sec:distortion}   

    \begin{figure*}[!ht]
       \centering
        \includegraphics[width=\hsize]{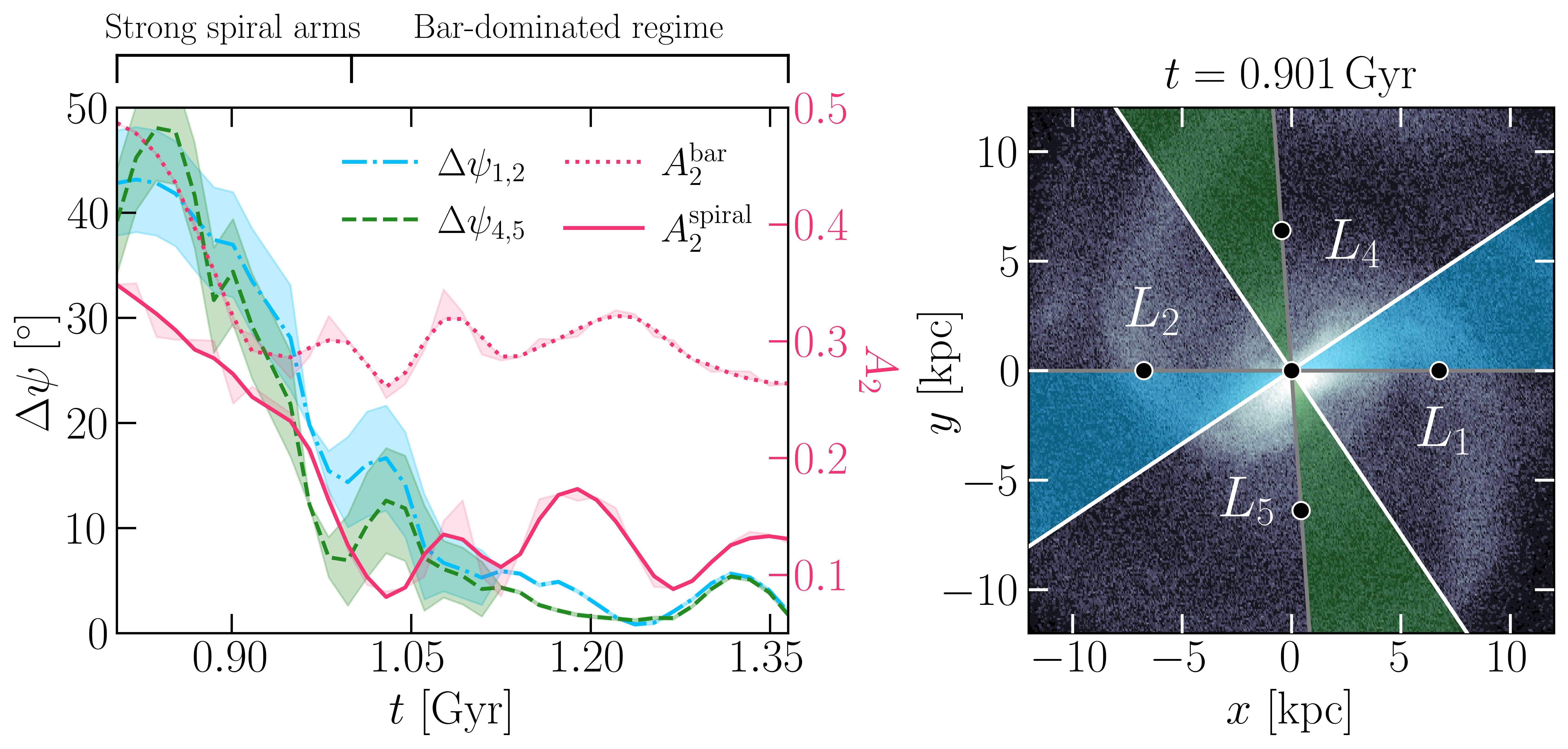}
            \caption{
            \textit{Left}: Time evolution of the angular offset of the equilibium points $L_1$ and $L_2$  with respect to the semimajor axis of the bar, $\Delta\psi_{1,2}$ (blue dotted-dashed line), and the angular displacement of $L_4$ ans $L_5$ with respect to the semi-minor axis of the bar, $\Delta\psi_{4,5}$ (green dashed line). The pink lines trace the evolution of the $m=2$ Fourier amplitude, $A_2$, of the bar (dotted line, averaged over [$R_0,\,R_1$] kpc) and of the spiral arms (solid line, averaged over [$R_0, 10$] kpc). In this context, $R_0$ and $R_1$ denote the inner and outer edges of the bar region as defined in \citet{Dehnen2022}. Shaded regions around each curve indicate errors and deviations with respect to the smoothed trends.
            \textit{Right}: Face-on surface density of the stellar component of the galaxy at $t=0.901$ Gyr after the initial conditions. The white lines indicate the bar semi-major and semi-minor axes, while the grey lines indicate the orientation of the equilibrium points, which are represented by black points. The shaded blue and green sectors respectively highlight the deviations $\Delta\psi_{1,2}$ and $\Delta\psi_{4,5}$ at this time.
            }
             \label{fig:psi}
    \end{figure*}
    
    Fig.~\ref{fig:psi} shows the impact of spiral arms on the gravitational potential of a barred galaxy.
    As long as the pattern speed of the bar does not vary significantly (see Fig.~1b of \citetalias{TST2026a}), a purely barred potential is expected to exhibit a stationary configuration in the rotating frame, in which the Lagrangian points $L_1$ and $L_2$ are aligned with the semi-major axis of the bar, while $L_4$ and $L_5$ lie symmetrically along its semi-minor axis \citep{BinneyTremaine}.
    However, the evolution of $\Delta\psi_{1,2}$ (the angular deviation of $L_1$ and $L_2$ with respect to the bar semi-major axis; blue dash-dotted curve) and $\Delta\psi_{4,5}$ (the angular deviation of $L_4$ and $L_5$ with respect to the semi-minor axis; green dashed curve), shown in the left panel of Fig.~\ref{fig:psi}, reveal systematic departures from this idealised configuration in the presence of spiral structure.\\
    
    Two different dynamical regimes can be identified from the temporal evolution of the $m=2$ Fourier amplitudes shown in the left panel of Fig.~\ref{fig:psi} (pink curves). 
    The pink solid curve traces the amplitude of the spiral arms, $A_2^\mathrm{spiral}$, averaged over the radial range $[R_1,10]\,\text{kpc}$; while the pink dotted curve corresponds to the bar component, $A_2^\mathrm{bar}$, averaged over $[R_0,R_1]\,\text{kpc}$.
    In this context, $R_0$ and $R_1$ respectively denote the inner and outer edges of the bar region as defined in Appendix B of \citet{Dehnen2022}.
    Although the bar remains dynamically dominant throughout the entire time interval ($A_2^\mathrm{bar} > 0.25$), the relative contribution of the spiral structure varies significantly with time.
    At early times ($t \lesssim 1.05$ Gyr), the spiral amplitude reaches relatively high values and becomes comparable to that of the bar, indicating the presence of strong, coherent bisymmetric arms. 
    In this regime, the spiral structure represents a significant non-axisymmetric contribution to the overall potential. 
    At later times ($t \gtrsim 1.05$~Gyr), however, $A_2^\mathrm{spiral}$ decreases substantially, while the bar amplitude remains approximately constant. 
    This marks a transition to a configuration in which the spiral arms are weaker and the potential is increasingly dominated by the bar.
    {Although this evolution naturally separates into phases with relatively strong and weak spiral arms, the displacement of the equilibrium points is itself continuous. No abrupt jumps between distinct configurations are observed; instead, the angular offsets evolve smoothly as the relative contribution of the spiral component gradually changes.}
    
    This variation in the relative strength of the spiral component is directly reflected in the behaviour of the angular deviations $\Delta\psi_{1,2}$ and $\Delta\psi_{4,5}$. 
    During the phase in which the spiral arms are strong ($t \lesssim 1.05$~Gyr), both quantities attain large values, with deviations reaching $20$–$50^\circ$. In contrast, once the spiral amplitude declines ($t \gtrsim 1.05$ Gyr), the angular offsets decrease significantly and remain below $10^\circ$.
    The close temporal correspondence between the evolution of $A_2^\mathrm{spiral}$ and the angular deviations indicates that the displacement of the equilibrium points is driven by the spiral component of the potential. 
    {This interpretation is consistent with both the nearly constant bar pattern speed over the same time interval (see Fig.~1 of \citetalias{TST2026a}) and the different timescales involved. While $\Delta\psi_{1,2}$ and $\Delta\psi_{4,5}$ vary over the $\sim200$~Myr period during which the strong spiral arms weaken, the bar evolves on its much longer secular timescale (\citealt{RocaFabrega2013}). 
    In addition, changes in the bar pattern speed would mainly shift the equilibrium points radially, whereas the observed evolution is dominated by smooth angular displacements that closely track $A_2^\mathrm{spiral}$. Together, these arguments indicate that the measured offsets are primarily induced by the spiral structure rather than by the secular evolution of the bar.}
    When the spiral structure is strong, it introduces a substantial non-axisymmetric contribution to the potential that displaces the Lagrangian points from their nominal locations along the bar major axis toward the spiral arms loci. 
    As the spiral weakens, this displacement diminishes, and the equilibrium points progressively recover the alignment expected for a purely barred potential.

    The right panel of Fig.~\ref{fig:psi} overlays the angular sectors $\Delta\psi_{1,2}$ and $\Delta\psi_{4,5}$ on the face-on surface density of the galaxy at $t = 0.901$ Gyr after the initial conditions, providing a visual illustration of the physical interpretation of the angular values in the left panel of Fig.~\ref{fig:psi}. 
    The blue shaded region represents the angular offset of the saddle points $L_1$–$L_2$ relative to the bar semi-major axis, while the green shaded region shows the corresponding deviation of the maxima $L_4$–$L_5$ with respect to the semi-minor axis.

    \begin{figure}[!t]
       \centering
       \includegraphics[width=\hsize]{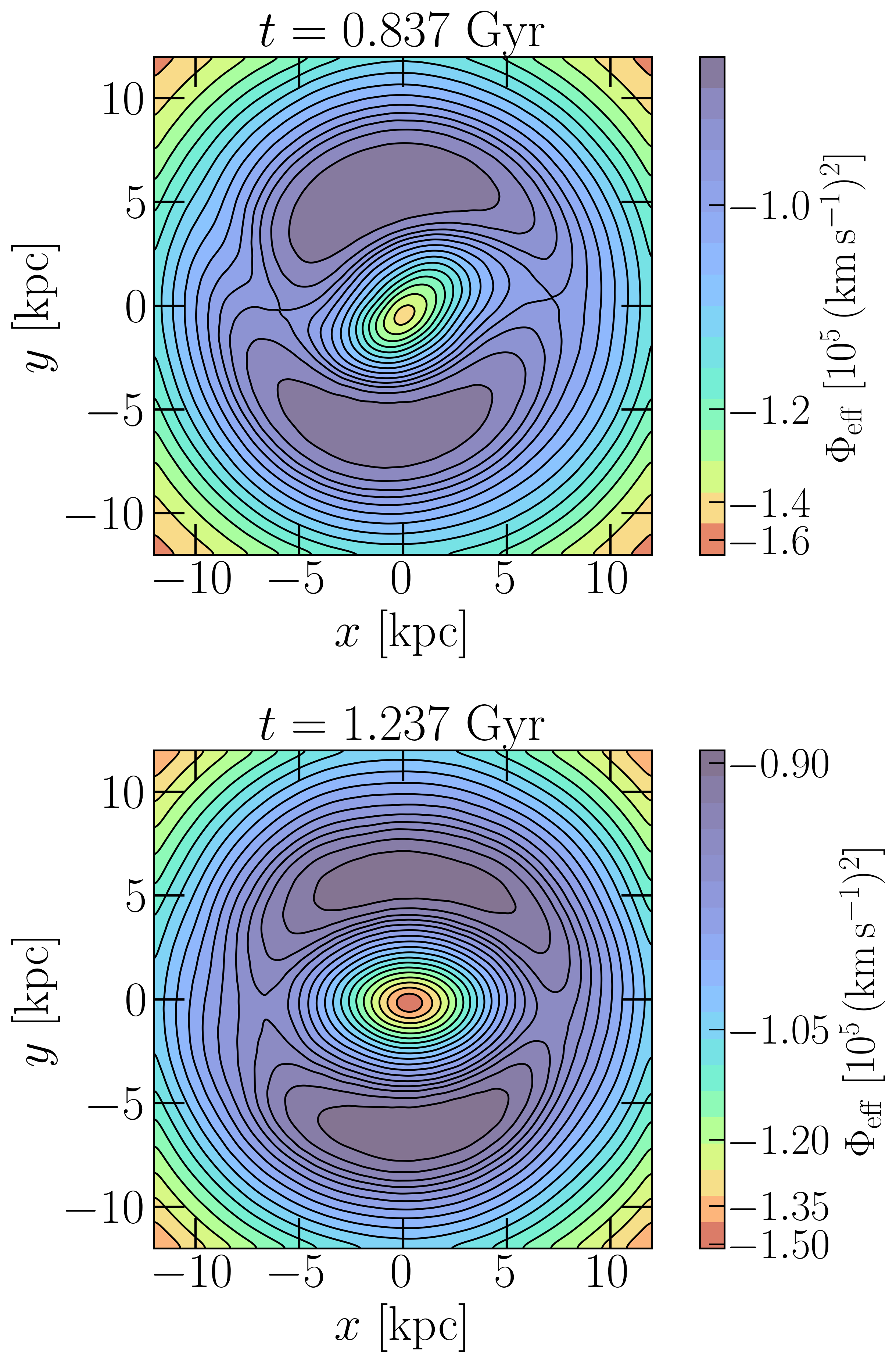}
        \caption{Effective gravitational potential $\Phi_\mathrm{eff}$ of the barred galaxy at two different stage: $t=0.837,\mathrm{Gyr}$ (top panel; strong spiral-arm regime) and $t=1.237,\mathrm{Gyr}$ (bottom panel; bar-dominated regime). Colours represent $\Phi_\mathrm{eff}$ values, with warmer tones indicating deeper effective potential wells and colder colours indicating higher effective potential values. Black lines denote equipotential contours.}
        \label{fig:poteffbrut}
    \end{figure}
    
    To further verify that these displacements are not an artefact introduced by the \texttt{AGAMA} reconstruction of the potential, Fig.~\ref{fig:poteffbrut} shows the effective gravitational potential $\Phi_\mathrm{eff}$ computed directly from the $N$-body particle distribution without assuming any symmetry or decomposition.
    For each snapshot, $10^5$ particles were randomly selected, and the gravitational potential was evaluated explicitly at their positions by summing the contribution of all particles in the simulation, including the dark matter halo component.
    The resulting effective potential was subsequently averaged on a regular mesh with $100\,\text{pc}\,\times100\,\text{pc}$ cells spanning a $20\,\text{kpc}\,\times20\,\text{kpc}$ grid, therefore providing a fully self-consistent brute-force estimate of the gravitational field.    
    The two panels of Fig.~\ref{fig:poteffbrut} correspond to the two dynamical regimes previously identified in Fig.~\ref{fig:psi}: a phase with strong, self-gravitating spiral arms ($t=0.837$ Gyr; upper panel), and a later bar-dominated phase in which the spiral structure has substantially weakened ($t=1.237$ Gyr; lower panel).
    Despite the absence of any imposed symmetry and the markedly different morphologies of the two snapshots, the equilibrium structure of the rotating-frame effective potential is clearly recognisable in both cases.
    The central minimum associated with $L_3$ is located at the galactic centre, while the banana-shaped equipotential regions surrounding $L_4$ and $L_5$ are visible around the local maxima of $\Phi_\mathrm{eff}$.
    The most significant difference between the two regimes concerns the location of the saddle equilibrium points $L_1$ and $L_2$.
    In the bar-dominated phase ($t=1.237$ Gyr; lower panel), these points are located near the ends of the bar, approximately along its major axis, in the regions where the equipotential contours pinch together, revealing the saddle-point structure of the effective potential.
    By contrast, during the strong spiral-arm phase ($t=0.837$ Gyr; upper panel), the same contour structure becomes noticeably distorted and misaligned with respect to the bar major axis.
    The saddle regions corresponding to $L_1$ and $L_2$ are therefore displaced away from the standard barred-galaxy configuration expected in a potential dominated by the galactic bar.

    Figure~\ref{fig:poteffbrut} thus provides direct visual confirmation that the equilibrium-point distribution depends sensitively on the strength of the spiral perturbation.
    When the spiral arms become sufficiently massive and self-gravitating, they introduce a non-axisymmetric perturbation to the rotating-frame potential, shifting the locations of the equilibrium points away from the principal axes of the bar.
    As the spiral structure weakens, the equipotential contours recover a much more symmetric morphology, and the equilibrium points progressively realign with the bar axes.
    This independently confirms that the angular displacements measured in Fig.~\ref{fig:psi} are genuine dynamical effects driven by the self-gravity of the spiral structure, rather than numerical artefacts associated with the potential reconstruction method.

    This behaviour is, in fact, consistent with previous theoretical and numerical studies of barred potentials.
    Departures of the equilibrium points from the principal bar axes have long been recognised as a natural consequence of non-axisymmetric distortions in the effective potential, either produced by offsets between the disc and bar centres of mass or by intrinsic asymmetries in the bar density distribution itself.
    Early works by \citet{deVaucouleursFreeman1970}, as well as by \citet{ColinAthanassoula1989}, already demonstrated that such asymmetries deform the contours of the effective isopotential and can substantially modify both the number and spatial configuration of equilibrium points in barred galaxies.
    More recently, \citet{SanchezMartin2023, SanchezMartin2026} showed explicitly that introducing bar offsets or asymmetric bar morphologies leads to measurable shifts in the equilibrium-point locations, together with significant distortions of the associated invariant manifolds.
    In this context, the offsets identified here should therefore be interpreted as the natural dynamical response of the system to the non-axisymmetric forcing generated by the self-gravitating spiral structure.
    Qualitative descriptions for these equilibrium-point deviations have previously been reported in numerical investigations of barred systems \citep[e.g.][]{Tsoutsis2009, Patsis2010, Athanassoula2012, Lokas2016, Efthymiopoulos2019}. 
    Building upon this picture, the present study provides a quantitative characterisation of this effect for the first time.

     \begin{figure*}[!ht]
       \centering
       \includegraphics[width=\hsize]{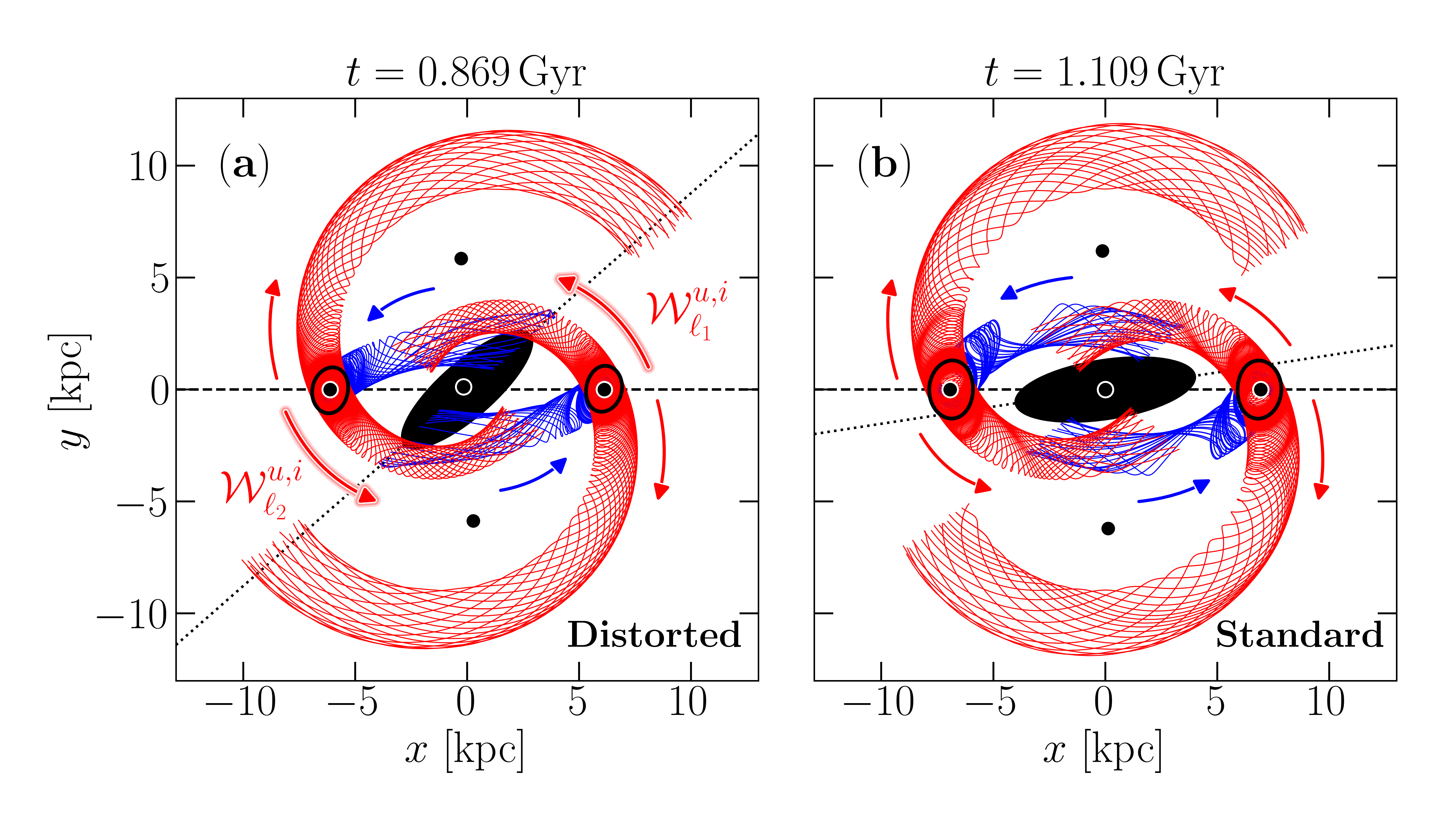}
            \caption{
            Schematic representation of the invariant manifold branches at $t = 0.869$ Gyr (left panel) and $t = 1.109$ Gyr (right panel). The black dots mark the positions of the Lagrangian points, $L_i$, with $i \in \{1,...,5\}$. The black closed curves around $L_1$ and $L_2$ represent the corresponding Lyapunov orbits. The red curves show the unstable invariant manifolds, $\mathcal{W}_{\ell_i}^u$, $i \in \{1,2\}$, emanating from these orbits, while the blue curves correspond to the inner stable manifolds, $\mathcal{W}_{\ell_i}^{s,i}$, which asymptotically approach them. The outer stable branches are omitted for clarity. Arrows indicate the direction of motion along each manifold branch. The black ellipse represents the bar, with a dotted line marking the orientation of its major axis. The dashed line indicates the alignment of the $L_1$–$L_2$ equilibrium points.
            }
             \label{fig:esquema}
    \end{figure*}

\section{Strengthening and weakening of the arms}\label{sec:weakening} 

\subsection{Standard and distorted manifold configurations}\label{sec:branches}

    Following the analysis presented in Sec.~\ref{sec:distortion} on the displacement of the equilibrium points produced by strong spiral arms, we now examine how the invariant manifolds associated with $L_1-L_2$ influence the evolution of the spiral structure when the saddle points are misaligned from the bar major axis.
    
    Figure~\ref{fig:esquema} illustrates the configuration of the invariant manifolds at two representative times. 
    Panel (a) corresponds to a representative snapshot in which the spiral arms are strong and self-gravitating ($t = 0.869$ Gyr), while panel (b) shows a later stage in which the bar dominates and the spiral structure is significantly weaker ($t = 1.109$ Gyr). 
    In both cases, each panel displays the branches of the invariant manifolds\footnote{The exterior stable branches of the invariant manifolds are omitted in this analysis (Fig.~\ref{fig:esquema}) for clarity.} (red and blue curves) together with the bar (black ellipse, constructed using $R_1$ from the DSS method \citep{Dehnen2022}), allowing for a direct comparison of their relative orientations.
    Importantly, spiral arms remain connected to the bar ends at all times, irrespective of their amplitude.
    
    In the weak-spiral regime (Fig.~\ref{fig:esquema}b), the Lagrangian points $L_1$ and $L_2$ are nearly aligned with the semi-major axis of the bar. 
    In this configuration, the invariant manifolds exhibit the standard morphology: the branches emanating from the Lyapunov orbits (black curves) originate close to the ends of the bar and naturally extend into well-defined structures, which in the case of the unstable exterior branches give rise to the spiral arms, as shown in \citetalias{TST2026a} and \citetalias{TST2026b}. 
    The resulting geometry favours a coherent connection between the bar and the spiral arms.

    By contrast, in the strong-spiral regime (Fig.~\ref{fig:esquema}a), the equilibrium points are clearly misaligned with respect to the bar major axis, as indicated by the offset between the dashed (Lagrangian points) and dotted (bar) lines. 
    This geometric decoupling has important dynamical consequences. In particular, the manifolds no longer emerge from the immediate vicinity of the bar ends, but instead originate further out, already embedded within the spiral structure.
    Consequently, the configuration of the manifold branches is significantly altered. 
    The inner unstable branches, which in the aligned case remain confined near the bar region, now overlap with the inner parts of the spiral arms.
    Since unstable {branches of the} manifolds govern motion away from the equilibrium points, this implies that particles guided by these tubes are driven along directions that oppose the flow associated with the outer unstable branches (see the red arrows in Fig.~\ref{fig:esquema}a).
    This inward-directed component, clearly visible in Fig. 6 of \citetalias{TST2026b}, introduces a counter-streaming motion within the inner regions of the arms, as further analysed in Sec. \ref{sec:inward_flows} (Fig.~\ref{fig:inward_flow}).
    
    As a result, the spiral structure is no longer supported by a single coherent flow, but by the superposition of competing (and opposed) manifold-guided streams. 
    This reduces the constructive alignment of trajectories along the arms, weakening their orbital support and leading to a decrease in their global amplitude.
    In this sense, strong spiral arms are responsible for displacing the equilibrium points, and these displacements feed back into the dynamics of the manifolds in a way that undermines the strength of the spiral structure.
    These results highlight the key role played by the relative orientation between the bar and the Lagrangian points in regulating spiral arm strength. 
    When both are aligned (Fig.~\ref{fig:esquema}b), the manifolds sustain coherent, well-organised flows through the arms. 
    However, when the spiral-induced distortion shifts the equilibrium points away from the bar major axis (Fig.~\ref{fig:esquema}a), the resulting reconfiguration of the manifolds disrupts this coherence, leading to a progressive weakening of the spiral pattern.
    In this regime, strong spiral arms can even decouple from the bar, forming detached outer segments, while new arms later emerge from the re-stabilised equilibrium points configuration.

\subsection{Inward flows along the arms as spiral weakeners}\label{sec:inward_flows}

    \begin{figure*}[!ht]
       \centering
       \includegraphics[width=\hsize]{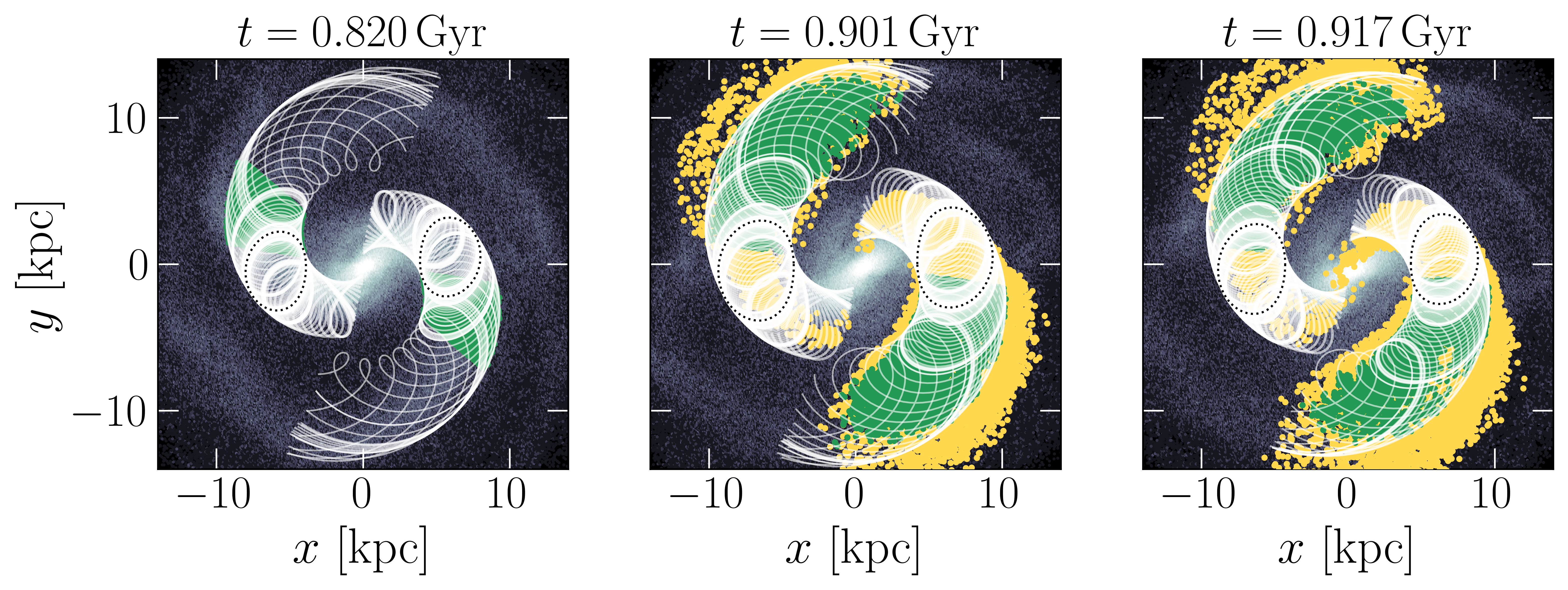}
        \caption{
        Time evolution of the bundle of particles initially trapped within the exterior unstable branches at $t=0.820$~Gyr, within the angular interval $10^\circ \leq \theta \leq 40^\circ$. A total of 26,617 particles are overlaid in each panel (13,967 associated with the $L_1$ exterior unstable branch and 12,650 with the $L_2$ branch). Subsequent panels follow the same particles forward in time, illustrating the evolution of their trajectories and trapping state, as defined in \citetalias{TST2026b}: green dots correspond to particles that remain trapped on the exterior branches (IN--IN) relative to the initial selection, while yellow dots indicate particles that escape from the trapped state (IN--OUT). White curves show the unstable branches of the invariant manifolds, including both the interior and exterior branches; the interior branches extend toward the bar region, whereas the exterior branches spread out into the spiral arms. The associated Lyapunov orbits are shown as black dotted curves. The background displays the face-on surface density of the disc, and each panel is labelled with the corresponding simulation time. {In this regime, the equilibrium points $L_1$ and $L_2$ are displaced from the bar semi-major axis (distorted configuration; see Fig.~\ref{fig:esquema}a). The outward flows still contribute to the spiral arm structure, but the inward flows now travel along the arms until reaching the bar, populating it directly.}
        }
        \label{fig:inward_flow}
    \end{figure*}

    In order to assess the impact of a distorted manifold configuration on the orbital behaviour of particles in the spiral arms\footnote{{Throughout this work, we consider that the spiral arms originate at the ends of the bar.}}, Fig.~\ref{fig:inward_flow} presents the time evolution of a bundle of particles initially trapped by the exterior unstable branches of the invariant manifolds within the angular interval $10^\circ \leq \theta \leq 40^\circ$, defined with respect to the position of the equilibrium points {$L_1$ and $L_2$}. 
    This selection isolates particles located along the spiral arms sufficiently distant from the bar region, yet still subject to the dynamical influence of the Lagrangian regions around $L_1$ and $L_2$.
    {Each of the particles in this bundle is tracked in the subsequent time steps, similarly to Fig.~6 of \citetalias{TST2026b}.
    Figure~\ref{fig:inward_flow} presents the evolution for the distorted manifold configuration (see Fig.~\ref{fig:esquema}a).
    The corresponding quantification of the trapping of particles by the unstable interior manifold branches, following the same methodology as in \citetalias{TST2026b}, is presented in Appendix~\ref{sec:appendix}.
    For comparison, the analogous evolution to Fig.~\ref{fig:inward_flow} in the standard manifold configuration (see Fig.~\ref{fig:esquema}b) is provided in Appendix~\ref{sec:no_inflow} (see Fig.~\ref{fig:no_inward_flow}).}
    
    In the standard configuration (described in Fig.~\ref{fig:esquema}b), spiral arms are predominantly populated by outward flows, as expected from the geometry of the exterior unstable branches. 
    These flows trace the spiral arms coherently, effectively supporting their structure. 
    Although inward flows are present along the interior unstable branches (inside co-rotation), they remain confined to the inner regions and primarily contribute to the inner ring, without directly interacting with the spiral structure —see Fig.~3 in \citet{Athanassoula2010}.
    Consequently, spiral arms are maintained through a well-organised, essentially one-way transport of material flowing along the manifolds {towards the outer disc}, preserving orbital coherence.

    \begin{figure*}[!ht]
       \centering
       \includegraphics[width=0.85\hsize]{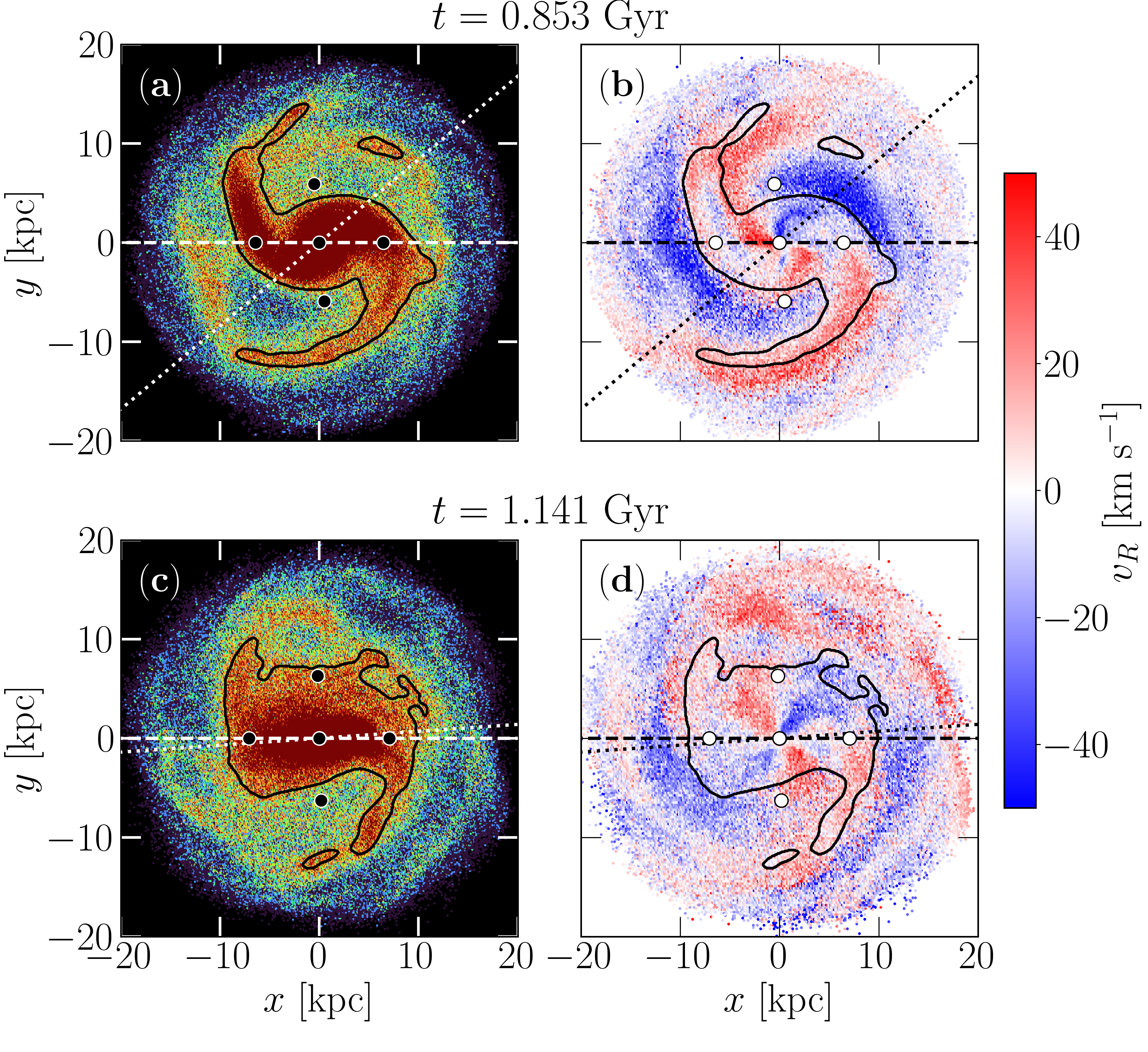}
        \caption{Face-on projection of the stellar component of the galaxy at two different times of the simulation. The top row (a,b) corresponds to a snapshot in the strong spiral-arm regime ($t = 0.853$ Gyr), while the bottom row (c,d) shows a later, bar-dominated stage ($t = 1.141$ Gyr). Left panels (a,c) display the surface density, and right panels (b,d) the corresponding $v_R$ maps. 
        The surface density is computed using hexagonal binning, with colours representing the projected stellar mass density on a logarithmic scale: warmer colours correspond to higher-density regions, while cooler colours trace lower-density ones.
        Black contours trace the main overdensities associated with the bar–spiral structure. The dots mark the positions of the Lagrangian points $L_1$ and $L_2$, while the dotted and dashed lines indicate the bar major axis and the $L_1$–$L_2$ direction, respectively.
        }
        \label{fig:comparison_vr}
    \end{figure*}
    
    By contrast, Fig.~\ref{fig:inward_flow}, which corresponds to the distorted manifold configuration (illustrated in Fig.~\ref{fig:esquema}a), reveals a qualitatively different behaviour. 
    In this case, as the equilibrium points are displaced by the strong spiral perturbation, 
    {the chaotic regions around $L_1$ and $L_2$ become stretched towards the spiral-arm overdensities, extending from the bar ends to these equilibrium points (see Fig.~\ref{fig:esquema}a). 
    As a result, inward and outward manifold flows co-exist throughout this region, so that a much larger fraction of the manifold-compatible population is able to access the interior branches than in the standard configuration}
    (see Fig.~\ref{fig:no_inward_flow}, $t=1.285$~Gyr). 
    The outward flows are mainly located along the outer portions of the arms, while the inward flows trace their inner regions, leading to the co-existence of particles with opposite radial motions at different azimuths of the arms. 
    Notably, the inward flow follows the spiral arms all the way back toward the bar region ($t=0.917$~Gyr, Fig.~\ref{fig:inward_flow}), directly repopulating the bar itself rather than accumulating in the surrounding ring, as occurs in the standard configuration (see~Fig. \ref{fig:no_inward_flow}). 
    Importantly, the inflows identified both in Fig.~\ref{fig:inward_flow} are channelled through the interior unstable branches of the invariant manifolds, in agreement with the schematic picture illustrated in Fig.~\ref{fig:esquema}.
    {Appendix~\ref{sec:appendix} presents a quantification of the manifold-trapping along the unstable interior branches during the strong spiral arm regime.}
    {The essential difference between the two manifold configurations is therefore not the existence of inward flows, but the context in which they take place. 
    In the standard configuration, they remain confined to the inner ring (see Fig.~\ref{fig:no_inward_flow}), whereas in the distorted configuration they affect the inner parts of the spiral arms themselves, providing a direct pathway that transports material from the arms back into the bar.}

    This interpretation is further supported by the kinematic information presented in Fig.~\ref{fig:comparison_vr}, which displays the spatial distribution of the galactocentric radial velocity, $v_R$, at two representative epochs.
    The top row corresponds to a phase with strongly developed spiral arms (Fig.~\ref{fig:comparison_vr}a,b), while the bottom row illustrates a later, bar-dominated stage (Fig.~\ref{fig:comparison_vr}c,d). The left panels display the surface density as seen face-on, whereas the right panels show the corresponding $v_R$ maps. In all cases, black contours trace the main overdensities associated with the bar–spiral structure, the dots mark the location of the Lagrangian points, and the dotted and dashed lines indicate the orientation of the bar and the $L_1$–$L_2$ axis, respectively.
    A clear correspondence emerges between the predicted manifold configurations and the velocity fields. In particular, the equilibrium points $L_1$ and $L_2$ systematically coincide with the $v_R = 0$ interface, delineating the transition between inflow and outflow regions. 
    Moreover, the expected radial flow pattern is robustly recovered: inside corotation (i.e. at smaller radii) the dominant motion is inward ($v_R < 0$), whereas outside corotation (i.e. at larger radii) particles predominantly exhibit outward motion ($v_R > 0$), as expected from the saddle behaviour of $L_1$ and $L_2$.
    Crucially, the distortions and asymmetries of the invariant manifold configurations are directly imprinted in the morphology of the velocity field.
    In the bar-dominated phase (panels c–d), the spiral arms are largely characterised by coherent outward radial flows. 
    In contrast, during the strong spiral phase (panels a–b), outward motions along the arms are mainly confined to regions beyond the equilibrium points $L_1$ and $L_2$, while the inner portions of the arms are dominated by inflows. 
    This close agreement confirms that the modifications of the manifold geometry induced by the self-gravity of the spiral arms have a direct and measurable kinematic counterpart.

\section{Discussion} \label{sec:discussion}
\subsection{Interpretation of the results}
    The results presented in Sec.~\ref{sec:distortion} and Sec.~\ref{sec:weakening} have two major dynamical implications.
    First, the co-existence of inward- and outward-moving particles along the arm features in the regime where the spiral arms are comparable to the bar leads to a loss of orbital coherence in the spiral structure.
    The arms are not supported by a coherent flux of manifold-compatible particles escaping along the exterior unstable manifolds, but instead present regions with opposing kinematics, which naturally leads to their progressive weakening.
    This mechanism provides a physical explanation to the weakening of the potential well generated by the self-gravity of trapped particles in the exterior unstable branches of invariant manifolds in Fig.~A1 of \citetalias{TST2026b}.
    
    {Second, Fig.~\ref{fig:inward_flow} shows that the inward flows along the inner parts of the spiral arms efficiently return material from the arms to the bar region, counteracting the net outward transport expected in the standard manifold configuration along the arms.}
    In this distorted regime, the manifolds do not simply act as channels expelling material from the inner disc; instead, they partially redirect spiral-arm material inward back to the bar, weakening the spiral structure and preventing bar depletion.
    {We emphasise that these manifold-guided orbits are not intended to constitute the dynamical backbone of the bar itself, which is expected to be supported primarily by periodic and quasi-periodic orbits inside co-rotation \citep{Contopoulos1980, Pfenninger1984b, ContopoulosGrosbol1989, Athanassoula1992, Skokos2002a, Skokos2002b, Sellwood&Wilkinson1993, BinneyTremaine}. 
    Rather, they replenish the bar region, where they contribute to the chaotic envelope of the bar (see Fig.~16 of \citet{Voglis2006a} and Fig.~5 of {\citet{Patsis2009}, published in} \citet{ContopoulosPatsis2009}), while remaining available for later spiral episodes.}
    {Such inward transport may also have implications for radial migration and chemical evolution, as it provides a mechanism capable of redistributing material from the spiral arms and inner disc into the bar region, potentially extending the chemical properties of the arms toward the central parts of the galaxy.}
    {Consequently, during phases of strong spiral activity, the system does not simply drain the bar and inner disc through manifold-transport as typically expected in the traditional invariant manifolds framework. 
    Instead, the co-existence of opposite flows along the arms allows the inner regions to retain a reservoir of particles capable of fuelling subsequent episodes of spiral growth. 
    The distorted manifold configuration is therefore dynamically unfavorable for maintaining indefinitely strong spiral arms, since it simultaneously reduces the manifold support of the spiral pattern and promotes the recycling of material toward the centre of the galaxy.
    This suggests that the transition between a regime of strong spiral arms and bar-dominated scenario may be naturally explained using manifold theory.}

    {Our results are also compatible with the observed enhancement of star formation at bar ends \citep[e.g.][]{Sheth2002,  Fraser-McKelvie2020a, Fraser-McKelvie2020b}. 
    Although our present study does not include gas dynamics or star formation, the interplay between the distorted manifold flows and the gaseous component represents an interesting avenue for future work, as it may provide additional insight into the coupling between spiral arms, bar-end star formation, and manifold-driven transport.}

    {In our simulation, this weakening process operates on a timescale of the order of $\sim200$~Myr.}
    {We stress, however, that this value is specific to the particular simulation analysed here, and should not be regarded as a universal timescale: the rate at which the spiral perturbation weakens is expected to depend on the relative strength of the bar and spiral components, the amount of self-gravitating material trapped in the manifold structure, and other properties of the underlying galactic potential. Variations in these parameters are likely to yield different weakening timescales in other systems.}
    
\subsection{Relation to previous works}

{To our knowledge, the mechanism presented in this work provides the first self-consistent explanation for the weakening of manifold-supported spiral arms in barred galaxies. 
It also accounts for the recurrent behaviour commonly observed in $N$-body simulations, whereby spiral arms initially emerge from the bar ends but appear to weaken, or even disconnect, from the bar as the spiral perturbation becomes sufficiently strong.
Previous studies have approached the recurrent nature of manifold-driven spirals from a different angle. 
{\citet{Contopoulos2009}, published in} \citet{ContopoulosPatsis2009}, proposed a qualitative scenario in which spiral recurrence arises from a continuous recycling of material between the two spiral arms (see their Fig.~14). 
Particles are transported from one side of the bar to the other along the exterior unstable branches of the invariant manifolds associated with the $L_1$ and $L_2$ Lagrangian points, allowing material from one arm to populate the opposite arm. 
While this recycling mechanism naturally accounts for the exchange of material between the two arms, it does not by itself explain the weakening of the global bisymmetric spiral pattern: because the material remains confined to the manifold tubes outside co-rotation, the mechanism primarily redistributes particles between the two arms rather than reducing the overall amount of material supporting the spiral structure as a whole.
In contrast, our mechanism weakens the spiral pattern by redirecting manifold-guided particles into the bar region, rather than simply recycling material between the two spiral arms.}

{A second aspect of our results concerns the displacement of the saddle-unstable equilibrium points away from the bar major axis when the spiral perturbation is sufficiently strong. Such displacements have previously been reported in self-consistent $N$-body simulations \citep{Tsoutsis2009, Athanassoula2012, Lokas2016, Efthymiopoulos2019}, but in those studies they were interpreted merely as a by-product of the evolving density distribution --and hence the evolving gravitational potential-- rather than examined as a mechanism that actively reshapes the morphology and dynamical role of the invariant manifolds. 
Similarly, manifold studies based on prescribed barred--spiral potentials have generally focused on the orbital structure and resulting morphology of spiral arms \citep[e.g.][]{Patsis2006, Patsis2009, TsigaridiPatsis2013, TsigaridiPatsis2015, PatsisTsigaridi2017, Efthymiopoulos2020, Efthymiopoulos2024}, without explicitly linking equilibrium-point displacement to the strength of the spiral perturbation. 
A particularly interesting study is \citet{Patsis2010}, who reconstructed the potential of the barred galaxy NGC~1300, which possesses very prominent, bisymmetric, grand-design spiral arms, and found the equilibrium points displaced from the bar major axis (see their Fig.~16).
Although this effect was not the focus of that work and its implications for the evolution of manifold-supported spiral structure were not explored, our results are compatible with their findings and provide a theoretical context in which they can be consistently interpreted.}

In the standard manifold picture, the saddle-unstable equilibrium points --and hence, by definition, co-rotation-- lie close to the bar ends. 
Since we show that sufficiently strong spiral perturbations displace these equilibrium points away from the bar major axis, the standard association between the bar ends and the saddle equilibrium points breaks down. 
Instead, $L_1$ and $L_2$ become embedded within the inner parts of the spiral arms.
{Our results therefore imply that the apparent connection between the ends of the bar and the onset of the spiral arms cannot always be used as a proxy for the location of the saddle equilibrium points or of co-rotation.}

{Finally, Appendix~\ref{sec:grecs} provides an extended discussion of our results in the context of the two complementary formulations of invariant-manifold theory. 
In particular, it clarifies how the present work, which adopts the flux-tube formulation, bears on the alternative apocentric-manifold perspective.}

\subsection{Caveats and limitations}
    In a strict sense, the direct applicability of the results presented in this work is formally restricted to systems sharing similar dynamical properties. 
    In the present set-up, the bar–spiral morphology corresponds to a configuration in which the spiral arms remain attached to the ends of the bar at all times and co-rotate at their connection, sharing a common pattern speed, as reported in \citet{RocaFabrega2013} (see Fig.~4 therein). 
    This implies a strong dynamical coupling between the bar and the inner spiral structure, consistent with a manifold-driven (or bar-driven) spiral scenario. 
    {In the $N$-body simulation analysed here, the spiral pattern speed outside the bar–spiral interface decreases only mildly with radius, indicating a coherent spiral pattern despite the absence of a perfectly rigid rotation. 
    In the calculations presented in this work, however, each individual snapshot is analysed assuming a single pattern speed at all radii corresponding to that of the bar.}
    
    Under these conditions, invariant manifolds provide a particularly suitable framework to interpret both the morphology and kinematics of the spiral structure. 
    However, in systems where the spiral arms are not anchored to the bar ends, or where multiple pattern speeds co-exist, the analysis becomes far more complex \citep[see][]{Efthymiopoulos2020, Efthymiopoulos2024}.
    In such cases, the trapping efficiency, as well as the balance between inward and outward radial flows, could differ significantly from the behaviour reported here.
    Nevertheless, the mechanisms identified in this work —namely, the distortion of manifold geometry by spiral self-gravity, the emergence of competing inward and outward flows, and the resulting loss of orbital coherence— are expected to remain qualitatively relevant in a broader context. 
    In this sense, while the quantitative results are specific to the adopted configuration, the underlying physical picture may apply more generally to barred galaxies with strongly coupled bar–spiral structures, and provides a useful framework for interpreting the interplay between spiral strength, radial migration, and bar evolution.

\section{Conclusions}\label{sec:conclusions}
In this work, we have studied the impact of spiral arms on the configuration of the Lagrangian points and explored the dynamical consequences for the associated invariant manifolds in an $N$-body simulation of an isolated barred galaxy.
The principal conclusions of this work can be summarised as follows:

\begin{itemize}
    \item[1.] We find that the presence of a strong spiral component induces significant angular displacements of the equilibrium points with respect to the bar axes, with deviations reaching up to $20$–$50^\circ$. These shifts are tightly correlated with the temporal evolution of the spiral amplitude, demonstrating that the spiral structure itself is responsible for perturbing the underlying potential away from the idealised barred configuration.
    This deviation from the Lagrangian points was qualitatively described in works such as \citet{Tsoutsis2009, Patsis2010, Athanassoula2012, Lokas2016}, and subsequently in \cite{Efthymiopoulos2019}, all of them showing that the spiral component of the gravitational potential exerts an additional force that is asymmetric with respect to the bar's semi-major axis, thereby shifting the position of the Lagrangian points away from it.
    In the present work, we provide the first quantitative measurement of this effect.\\
    
    \item[2.] The misalignment of the equilibrium points has important implications for the manifold dynamics. 
    When the Lagrangian points are nearly aligned with the bar, the invariant manifolds support spiral arms with a well-defined outward flow along the outer unstable branches, whereby particles travel from the inner to the outer parts of the arms, in agreement with previous studies \citep[e.g.][]{MRG2006, MRG2007, Athanassoula2009b, Athanassoula2009a, Voglis2006a, Tsoutsis2008, Athanassoula2010, Athanassoula2012, Baba2015, Lokas2016, Efthymiopoulos2019, Efthymiopoulos2020, Zouloumi2024}. 
    However, when strong spiral arms displace the equilibrium points, the geometry of the manifolds is substantially altered with respect to the standard bar-spiral configuration. 
    In this regime, the manifold branches no longer emerge from the bar ends in a coherent manner, and competing flows develop, leading to a depletion of material {in} the inner parts of the arms {towards the bar region. This progressively weakens the spiral pattern and may ultimately lead to a break in the bar--spiral connection}.
    This interpretation also helps to place our previous results in context: in \citetalias{TST2026b} the analysis was restricted to the outer unstable branch, although invariant manifolds correspond to the full set of associated tubes. 
    By including here both the inner and outer unstable branches, we obtain a more complete dynamical picture of the {recurrent} evolution of spiral arms.\\

    \item[3.] Our results suggest a self-regulating mechanism for spiral structure in barred galaxies. 
    While invariant manifolds naturally give rise to spiral arms in barred galaxies, the growth of the spiral structure modifies the gravitational potential in such a way that the equilibrium points become misaligned, which in turn disrupts the manifold-supported flows that sustain them. 
    As the spiral weakens, the system tends to recover the standard barred-galaxy configuration, allowing the manifolds to once again support the strengthening of spiral arms.
    In this sense, invariant manifolds do not only provide a framework to explain the formation of spiral structure, but also its recurrent weakening and reformation, pointing to a cyclic, self-regulated evolution of spiral arms in barred galaxies.
    {This mechanism provides a natural dynamical explanation for the recurrent weakening and reformation of spiral arms found in numerical simulations of barred galaxies, suggesting that strong spiral arms are unlikely to remain in a steady, high-amplitude state over indefinitely long timescales.}\\

    \item[4.] The findings in this work also provide compelling evidence that the bar is not indefinitely depleted by the manifolds, but instead participates in a sustained and self-consistent exchange of material (not only with the disc, but also) with the spiral arms.
    While the contribution of the stable exterior branches is not considered here—since no significant overdensity is observed in their projection onto configuration space for bisymmetric, grand-design spirals—, we find that the spiral-arm-induced misalignments of the Lagrangian points play a key role in regulating this exchange. 
    In particular, these misalignments weaken the outward manifold-supported flows and promote the return of material from the spiral arms back into the bar region through the inner unstable branches. 
    This mechanism establishes a self-regulated cycle that redistributes mass without leading to any long-term depletion of the bar.
\end{itemize}

\begin{acknowledgements}
      T.S.T. acknowledges that this work was partially supported by the FI-STEP predoctoral grant program of the Department of Research and Universities of the Generalitat de Catalunya, co-funded by the European Social Fund Plus (reference number: 2025 STEP 00196). 
      M.R.G. and T.S.T acknowledge that this work was (partially) supported by the Spanish MICIN/AEI/10.13039/501100011033 and by "ERDF A way of making Europe" by the European Union through grants PID2021-122842OB-C21 and PID2024-157964OB-C21, the Institute of Cosmos Sciences University of Barcelona (ICCUB, Unidad de Excelencia María de Maeztu) through grant CEX2024-001451-M and the project 2021-SGR-00679 GRC de l’Agència de Gestió d’Ajuts Universitaris i de Recerca (Generalitat de Catalunya). 
      S.R.F. acknowledges the finantial support by the Swedish National Space Agency Senior career grant with number 2025-00181, and the Spanish Ministry of Science and Innovation through the research grants: PID2021-123417OBI00, funded by MCIN/AEI/10.13039/501100011033/FEDER, EU; PCI2022-135023-2, funded by MCIN/AEI/10.13039/501100011033 and the EU “NextGenerationEU” / PRTR; and PID2024-157374OBI00, funded by MI-CIU/AEI/10.13039/501100011033/FEDER, EU.
\end{acknowledgements}

\bibliographystyle{aa}
\bibliography{bibliography}

@ARTICLE{Athanassoula1992,
       author = {{Athanassoula}, E.},
        title = "{Morphology of bar orbits.}",
      journal = {\mnras},
         year = 1992,
        month = nov,
       volume = {259},
        pages = {328-344},
          doi = {10.1093/mnras/259.2.328},
       adsurl = {https://ui.adsabs.harvard.edu/abs/1992MNRAS.259..328A}
}

@ARTICLE{Athanassoula2009a,
       author = {{Athanassoula}, E. and {Romero-G{\'o}mez}, M. and {Masdemont}, J.~J.},
        title = "{Rings and spirals in barred galaxies - I. Building blocks}",
      journal = {\mnras},
         year = 2009,
        month = mar,
       volume = {394},
       number = {1},
        pages = {67-81},
          doi = {10.1111/j.1365-2966.2008.14273.x},
archivePrefix = {arXiv},
       eprint = {0811.4056},
 primaryClass = {astro-ph},
       adsurl = {https://ui.adsabs.harvard.edu/abs/2009MNRAS.394...67A}
}

@ARTICLE{Athanassoula2009b,
       author = {{Athanassoula}, E. and {Romero-G{\'o}mez}, M. and {Bosma}, A. and {Masdemont}, J.~J.},
        title = "{Rings and spirals in barred galaxies - II. Ring and spiral morphology}",
      journal = {\mnras},
         year = 2009,
        month = dec,
       volume = {400},
       number = {4},
        pages = {1706-1720},
          doi = {10.1111/j.1365-2966.2009.15583.x},
archivePrefix = {arXiv},
       eprint = {0910.0757},
 primaryClass = {astro-ph.CO},
       adsurl = {https://ui.adsabs.harvard.edu/abs/2009MNRAS.400.1706A}
}

@ARTICLE{Athanassoula2010,
       author = {{Athanassoula}, E. and {Romero-G{\'o}mez}, M. and {Bosma}, A. and {Masdemont}, J.~J.},
        title = "{Rings and spirals in barred galaxies - III. Further comparisons and links to observations}",
      journal = {\mnras},
         year = 2010,
        month = sep,
       volume = {407},
       number = {3},
        pages = {1433-1448},
          doi = {10.1111/j.1365-2966.2010.17010.x},
archivePrefix = {arXiv},
       eprint = {1005.2943},
 primaryClass = {astro-ph.CO},
       adsurl = {https://ui.adsabs.harvard.edu/abs/2010MNRAS.407.1433A}
}

@ARTICLE{Athanassoula2012,
       author = {{Athanassoula}, E.},
        title = "{Manifold-driven spirals in N-body barred galaxy simulations}",
      journal = {\mnras},
         year = 2012,
        month = oct,
       volume = {426},
       number = {1},
        pages = {L46-L50},
          doi = {10.1111/j.1745-3933.2012.01320.x},
archivePrefix = {arXiv},
       eprint = {1207.4590},
 primaryClass = {astro-ph.GA},
       adsurl = {https://ui.adsabs.harvard.edu/abs/2012MNRAS.426L..46A}
}

@ARTICLE{Baba2013,
       author = {{Baba}, Junichi and {Saitoh}, Takayuki R. and {Wada}, Keiichi},
        title = "{Dynamics of Non-steady Spiral Arms in Disk Galaxies}",
      journal = {\apj},
         year = 2013,
        month = jan,
       volume = {763},
       number = {1},
          eid = {46},
        pages = {46},
          doi = {10.1088/0004-637X/763/1/46},
archivePrefix = {arXiv},
       eprint = {1211.5401},
 primaryClass = {astro-ph.GA},
       adsurl = {https://ui.adsabs.harvard.edu/abs/2013ApJ...763...46B}
}

@ARTICLE{Baba2015,
       author = {{Baba}, Junichi},
        title = "{Short-term dynamical evolution of grand-design spirals in barred galaxies}",
      journal = {\mnras},
         year = 2015,
        month = dec,
       volume = {454},
       number = {3},
        pages = {2954-2964},
          doi = {10.1093/mnras/stv2220},
archivePrefix = {arXiv},
       eprint = {1509.07239},
 primaryClass = {astro-ph.GA},
       adsurl = {https://ui.adsabs.harvard.edu/abs/2015MNRAS.454.2954B}
}

@BOOK{BinneyTremaine,
       author = {{Binney}, James and {Tremaine}, Scott},
        title = "{Galactic Dynamics: Second Edition}",
         year = 2008,
       adsurl = {https://ui.adsabs.harvard.edu/abs/2008gady.book.....B}
}

@ARTICLE{Buta2015,
       author = {{Buta}, Ronald J. and {Sheth}, Kartik and {Athanassoula}, E. and {Bosma}, A. and {Knapen}, Johan H. and {Laurikainen}, Eija and {Salo}, Heikki and {Elmegreen}, Debra and {Ho}, Luis C. and {Zaritsky}, Dennis and {Courtois}, Helene and {Hinz}, Joannah L. and {Mu{\~n}oz-Mateos}, Juan-Carlos and {Kim}, Taehyun and {Regan}, Michael W. and {Gadotti}, Dimitri A. and {Gil de Paz}, Armando and {Laine}, Jarkko and {Men{\'e}ndez-Delmestre}, Kar{\'\i}n and {Comer{\'o}n}, S{\'e}bastien and {Erroz Ferrer}, Santiago and {Seibert}, Mark and {Mizusawa}, Trisha and {Holwerda}, Benne and {Madore}, Barry F.},
        title = "{A Classical Morphological Analysis of Galaxies in the Spitzer Survey of Stellar Structure in Galaxies (S4G)}",
      journal = {\apjs},
         year = 2015,
        month = apr,
       volume = {217},
       number = {2},
          eid = {32},
        pages = {32},
          doi = {10.1088/0067-0049/217/2/32},
archivePrefix = {arXiv},
       eprint = {1501.00454},
 primaryClass = {astro-ph.GA},
       adsurl = {https://ui.adsabs.harvard.edu/abs/2015ApJS..217...32B}
}

@ARTICLE{CarlbergFreedman1985,
       author = {{Carlberg}, R.~G. and {Freedman}, W.~L.},
        title = "{Dissipative models of spiral galaxies}",
      journal = {\apj},
         year = 1985,
        month = nov,
       volume = {298},
        pages = {486-492},
          doi = {10.1086/163634},
       adsurl = {https://ui.adsabs.harvard.edu/abs/1985ApJ...298..486C}
}

@ARTICLE{ColinAthanassoula1989,
       author = {{Colin}, J. and {Athanassoula}, E.},
        title = "{Asymmetrical barred galaxies.}",
      journal = {\aap},
         year = 1989,
        month = apr,
       volume = {214},
        pages = {99-105},
       adsurl = {https://ui.adsabs.harvard.edu/abs/1989A&A...214...99C}
}

@ARTICLE{Contopoulos1980,
       author = {{Contopoulos}, G. and {Papayannopoulos}, Th.},
      journal = {\aap},
         year = 1980,
       volume = {92},
        pages = {33-46}
}

@ARTICLE{ContopoulosGrosbol1989,
       author = {{Contopoulos}, G. and {Grosbol}, P.},
        title = "{Orbits in barred galaxies}",
      journal = {\aapr},
         year = 1989,
        month = nov,
       volume = {1},
       number = {3-4},
        pages = {261-289},
          doi = {10.1007/BF00873080},
       adsurl = {https://ui.adsabs.harvard.edu/abs/1989A&ARv...1..261C}
}

@BOOK{Contopoulos2009,
  author    = {Contopoulos, G.},
  title     = {Chaos in Astronomy},
  publisher = {Springer},
  year      = {2009},
  editor    = {Contopoulos, G. and Patsis, P. A.},
  series    = {Astrophysics and Space Science Proceedings},
  volume    = {8},
  doi       = {10.1007/978-3-540-75826-6}
}

@BOOK{ContopoulosPatsis2009,
  editor    = {G. Contopoulos and P. A. Patsis},
  title     = {Chaos in Astronomy},
  series    = {Astrophysics and Space Science Proceedings},
  publisher = {Springer},
  address   = {Berlin and Heidelberg},
  year      = {2009},
  doi       = {10.1007/978-3-540-75826-6},
  isbn      = {978-3-540-75826-6}
}

@INPROCEEDINGS{deVaucouleursFreeman1970,
       author = {{de Vaucouleurs}, G. and {Freeman}, K.~C.},
        title = "{Structure and Dynamics of Barred Spiral Galaxies with an Asymmetric Mass Distribution}",
    booktitle = {The Spiral Structure of our Galaxy},
         year = 1970,
       editor = {{Becker}, Wilhelm and {Kontopoulos}, Georgios Ioannou},
       series = {IAU Symposium},
       volume = {38},
        month = jan,
        pages = {356},
       adsurl = {https://ui.adsabs.harvard.edu/abs/1970IAUS...38..356D}
}

@ARTICLE{Dehnen2022,
       author = {{Dehnen}, Walter and {Semczuk}, Marcin and {Sch{\"o}nrich}, Ralph},
        title = "{Measuring bar pattern speeds from single simulation snapshots}",
      journal = {\mnras},
         year = 2023,
        month = jan,
       volume = {518},
       number = {2},
        pages = {2712-2718},
          doi = {10.1093/mnras/stac3184},
archivePrefix = {arXiv},
       eprint = {2211.00674},
 primaryClass = {astro-ph.GA},
       adsurl = {https://ui.adsabs.harvard.edu/abs/2023MNRAS.518.2712D}
}

@ARTICLE{DiazGarcia2019,
       author = {{D{\'\i}az-Garc{\'\i}a}, S. and {Salo}, H. and {Knapen}, J.~H. and {Herrera-Endoqui}, M.},
        title = "{The shapes of spiral arms in the S$^{4}$G survey and their connection with stellar bars}",
      journal = {\aap},
         year = 2019,
        month = nov,
       volume = {631},
          eid = {A94},
        pages = {A94},
          doi = {10.1051/0004-6361/201936000},
archivePrefix = {arXiv},
       eprint = {1908.04246},
 primaryClass = {astro-ph.GA},
       adsurl = {https://ui.adsabs.harvard.edu/abs/2019A&A...631A..94D}
}

@ARTICLE{Efthymiopoulos2010,
       author = {{Efthymiopoulos}, C.},
        title = "{Special features of galactic dynamics: Disc dynamics}",
      journal = {European Physical Journal Special Topics},
         year = 2010,
        month = sep,
       volume = {186},
       number = {1},
        pages = {91-122},
          doi = {10.1140/epjst/e2010-01261-8},
       adsurl = {https://ui.adsabs.harvard.edu/abs/2010EPJST.186...91E}
}

@ARTICLE{Efthymiopoulos2019,
       author = {{Efthymiopoulos}, C. and {Kyziropoulos}, P.~E. and {P{\'a}ez}, R.~I. and {Zouloumi}, K. and {Gravvanis}, G.~A.},
        title = "{Manifold spirals, disc-halo interactions, and the secular evolution in N-body models of barred galaxies}",
      journal = {\mnras},
         year = 2019,
        month = apr,
       volume = {484},
       number = {2},
        pages = {1487-1505},
          doi = {10.1093/mnras/stz035},
archivePrefix = {arXiv},
       eprint = {1901.00692},
 primaryClass = {astro-ph.GA},
       adsurl = {https://ui.adsabs.harvard.edu/abs/2019MNRAS.484.1487E}
}

@ARTICLE{Efthymiopoulos2020,
       author = {{Efthymiopoulos}, C. and {Harsoula}, M. and {Contopoulos}, G.},
        title = "{Manifold spirals in barred galaxies with multiple pattern speeds}",
      journal = {\aap},
         year = 2020,
        month = apr,
       volume = {636},
          eid = {A44},
        pages = {A44},
          doi = {10.1051/0004-6361/201936871},
archivePrefix = {arXiv},
       eprint = {1910.06653},
 primaryClass = {nlin.CD},
       adsurl = {https://ui.adsabs.harvard.edu/abs/2020A&A...636A..44E}
}

@INPROCEEDINGS{Efthymiopoulos2024,
       author = {{Efthymiopoulos}, Christos and {Zouloumi}, Konstantina and {Harsoula}, Maria},
        title = "{Manifold spirals in barred-spiral galaxies with two pattern speeds: N-body and Milky Way models}",
    booktitle = {EAS2024, European Astronomical Society Annual Meeting},
         year = 2024,
        month = jul,
          eid = {292},
        pages = {292},
       adsurl = {https://ui.adsabs.harvard.edu/abs/2024eas..conf..292E}
}

@ARTICLE{Font2019,
       author = {{Font}, Joan and {Beckman}, John E. and {James}, Phil A. and {Patsis}, Panos A.},
        title = "{Spiral structure in barred galaxies. Observational constraints to spiral arm formation mechanisms}",
      journal = {\mnras},
         year = 2019,
        month = feb,
       volume = {482},
       number = {4},
        pages = {5362-5378},
          doi = {10.1093/mnras/sty2983},
archivePrefix = {arXiv},
       eprint = {1901.04725},
 primaryClass = {astro-ph.GA},
       adsurl = {https://ui.adsabs.harvard.edu/abs/2019MNRAS.482.5362F}
}

@ARTICLE{Fraser-McKelvie2020a,
       author = {{Fraser-McKelvie}, Amelia and {Arag{\'o}n-Salamanca}, Alfonso and {Merrifield}, Michael and {Masters}, Karen and {Nair}, Preethi and {Emsellem}, Eric and {Kraljic}, Katarina and {Krishnarao}, Dhanesh and {Andrews}, Brett H. and {Drory}, Niv and {Neumann}, Justus},
        title = "{SDSS-IV MaNGA: spatially resolved star formation in barred galaxies}",
      journal = {\mnras},
         year = 2020,
        month = jul,
       volume = {495},
       number = {4},
        pages = {4158-4169},
          doi = {10.1093/mnras/staa1416},
archivePrefix = {arXiv},
       eprint = {2005.08987},
 primaryClass = {astro-ph.GA},
       adsurl = {https://ui.adsabs.harvard.edu/abs/2020MNRAS.495.4158F}
}

@ARTICLE{Fraser-McKelvie2020b,
       author = {{Fraser-McKelvie}, Amelia and {Merrifield}, Michael and {Arag{\'o}n-Salamanca}, Alfonso and {Peterken}, Thomas and {Kraljic}, Katarina and {Masters}, Karen and {Stark}, David and {Fragkoudi}, Francesca and {Smethurst}, Rebecca and {Boardman}, Nicholas Fraser and {Drory}, Niv and {Lane}, Richard R.},
        title = "{SDSS-IV MaNGA: The link between bars and the early cessation of star formation in spiral galaxies}",
      journal = {\mnras},
         year = 2020,
        month = nov,
       volume = {499},
       number = {1},
        pages = {1116-1125},
          doi = {10.1093/mnras/staa2866},
archivePrefix = {arXiv},
       eprint = {2009.07859},
 primaryClass = {astro-ph.GA},
       adsurl = {https://ui.adsabs.harvard.edu/abs/2020MNRAS.499.1116F}
}

@ARTICLE{Grand2012b,
       author = {{Grand}, Robert J.~J. and {Kawata}, Daisuke and {Cropper}, Mark},
        title = "{The dynamics of stars around spiral arms}",
      journal = {\mnras},
         year = 2012,
        month = apr,
       volume = {421},
       number = {2},
        pages = {1529-1538},
          doi = {10.1111/j.1365-2966.2012.20411.x},
archivePrefix = {arXiv},
       eprint = {1112.0019},
 primaryClass = {astro-ph.GA},
       adsurl = {https://ui.adsabs.harvard.edu/abs/2012MNRAS.421.1529G}
}

@ARTICLE{Harsoula2016,
       author = {{Harsoula}, M. and {Efthymiopoulos}, C. and {Contopoulos}, G.},
        title = "{Analytical forms of chaotic spiral arms}",
      journal = {\mnras},
         year = 2016,
        month = jul,
       volume = {459},
       number = {4},
        pages = {3419-3431},
          doi = {10.1093/mnras/stw748},
archivePrefix = {arXiv},
       eprint = {1603.09151},
 primaryClass = {nlin.CD},
       adsurl = {https://ui.adsabs.harvard.edu/abs/2016MNRAS.459.3419H}
}

@ARTICLE{HarsoulaKatsanikas2025,
       author = {{Harsoula}, M. and {Katsanikas}, M.},
        title = "{The building blocks of 3D models of barred galaxies: I. The case of ring galaxies}",
      journal = {\aap},
         year = 2025,
        month = jul,
       volume = {699},
          eid = {A13},
        pages = {A13},
          doi = {10.1051/0004-6361/202554584},
       adsurl = {https://ui.adsabs.harvard.edu/abs/2025A&A...699A..13H}
}

@ARTICLE{KatsanikasPatsis2022,
       author = {{Katsanikas}, M. and {Patsis}, P.~A.},
        title = "{The phase space structure in the vicinity of vertical Lyapunov orbits around L$_{1,2}$ in a barred galaxy model}",
      journal = {\mnras},
         year = 2022,
        month = nov,
       volume = {516},
       number = {4},
        pages = {5232-5243},
          doi = {10.1093/mnras/stac2632},
archivePrefix = {arXiv},
       eprint = {2209.10249},
 primaryClass = {astro-ph.GA},
       adsurl = {https://ui.adsabs.harvard.edu/abs/2022MNRAS.516.5232K}
}

@ARTICLE{Koon2000,
       author = {{Koon}, Wang Sang and {Lo}, Martin W. and {Marsden}, Jerrold E. and {Ross}, Shane D.},
        title = "{Heteroclinic connections between periodic orbits and resonance transitions in celestial mechanics}",
      journal = {Chaos},
         year = 2000,
        month = jun,
       volume = {10},
       number = {2},
        pages = {427-469},
          doi = {10.1063/1.166509},
       adsurl = {https://ui.adsabs.harvard.edu/abs/2000Chaos..10..427K}
}

@ARTICLE{Lee2023,
       author = {{Lee}, Janice C. and {Sandstrom}, Karin M. and {Leroy}, Adam K. and {Thilker}, David A. and {Schinnerer}, Eva and {Rosolowsky}, Erik and {Larson}, Kirsten L. and {Egorov}, Oleg V. and {Williams}, Thomas G. and {Schmidt}, Judy and {Emsellem}, Eric and {Anand}, Gagandeep S. and {Barnes}, Ashley T. and {Belfiore}, Francesco and {Be{\v{s}}li{\'c}}, Ivana and {Bigiel}, Frank and {Blanc}, Guillermo A. and {Bolatto}, Alberto D. and {Boquien}, M{\'e}d{\'e}ric and {den Brok}, Jakob and {Cao}, Yixian and {Chandar}, Rupali and {Chastenet}, J{\'e}r{\'e}my and {Chevance}, M{\'e}lanie and {Chiang}, I-Da and {Congiu}, Enrico and {Dale}, Daniel A. and {Deger}, Sinan and {Eibensteiner}, Cosima and {Faesi}, Christopher M. and {Glover}, Simon C.~O. and {Grasha}, Kathryn and {Groves}, Brent and {Hassani}, Hamid and {Henny}, Kiana F. and {Henshaw}, Jonathan D. and {Hoyer}, Nils and {Hughes}, Annie and {Jeffreson}, Sarah and {Jim{\'e}nez-Donaire}, Mar{\'\i}a J. and {Kim}, Jaeyeon and {Kim}, Hwihyun and {Klessen}, Ralf S. and {Koch}, Eric W. and {Kreckel}, Kathryn and {Kruijssen}, J.~M. Diederik and {Li}, Jing and {Liu}, Daizhong and {Lopez}, Laura A. and {Maschmann}, Daniel and {Chen}, Ness Mayker and {Meidt}, Sharon E. and {Murphy}, Eric J. and {Neumann}, Justus and {Neumayer}, Nadine and {Pan}, Hsi-An and {Pessa}, Ismael and {Pety}, J{\'e}r{\^o}me and {Querejeta}, Miguel and {Pinna}, Francesca and {Rodr{\'\i}guez}, M. Jimena and {Saito}, Toshiki and {S{\'a}nchez-Bl{\'a}zquez}, Patricia and {Santoro}, Francesco and {Sardone}, Amy and {Smith}, Rowan J. and {Sormani}, Mattia C. and {Scheuermann}, Fabian and {Stuber}, Sophia K. and {Sutter}, Jessica and {Sun}, Jiayi and {Teng}, Yu-Hsuan and {Tre{\ss}}, Robin G. and {Usero}, Antonio and {Watkins}, Elizabeth J. and {Whitmore}, Bradley C. and {Razza}, Alessandro},
        title = "{The PHANGS-JWST Treasury Survey: Star Formation, Feedback, and Dust Physics at High Angular Resolution in Nearby GalaxieS}",
      journal = {\apjl},
         year = 2023,
        month = feb,
       volume = {944},
       number = {2},
          eid = {L17},
        pages = {L17},
          doi = {10.3847/2041-8213/acaaae},
archivePrefix = {arXiv},
       eprint = {2212.02667},
 primaryClass = {astro-ph.GA},
       adsurl = {https://ui.adsabs.harvard.edu/abs/2023ApJ...944L..17L}
}

@ARTICLE{Lokas2016,
       author = {{{\L}okas}, Ewa L.},
        title = "{Damping of the Milky Way Bar by Manifold-driven Spirals}",
      journal = {\apjl},
         year = 2016,
        month = oct,
       volume = {830},
       number = {1},
          eid = {L20},
        pages = {L20},
          doi = {10.3847/2041-8205/830/1/L20},
archivePrefix = {arXiv},
       eprint = {1607.08339},
 primaryClass = {astro-ph.GA},
       adsurl = {https://ui.adsabs.harvard.edu/abs/2016ApJ...830L..20L}
}

@ARTICLE{Meidt2025,
       author = {{van der Wel}, Arjen and {Meidt}, Sharon E.},
        title = "{Transforming Galaxies with EASE: Widespread structural changes enabled by short-lived spirals}",
      journal = {\aap},
         year = 2025,
        month = dec,
       volume = {704},
          eid = {A147},
        pages = {A147},
          doi = {10.1051/0004-6361/202556758},
archivePrefix = {arXiv},
       eprint = {2509.02847},
 primaryClass = {astro-ph.GA},
       adsurl = {https://ui.adsabs.harvard.edu/abs/2025A&A...704A.147V}
}

@ARTICLE{Meidt2026,
       author = {{Meidt}, Sharon E. and {van der Wel}, Arjen},
        title = "{Nexae in caverna: the secular evolution of disks via collectively excited, transient spiral structure}",
      journal = {arXiv e-prints},
         year = 2026,
        month = apr,
          eid = {arXiv:2604.04827},
        pages = {arXiv:2604.04827},
          doi = {10.48550/arXiv.2604.04827},
archivePrefix = {arXiv},
       eprint = {2604.04827},
 primaryClass = {astro-ph.GA},
       adsurl = {https://ui.adsabs.harvard.edu/abs/2026arXiv260404827M}
}

@ARTICLE{OllePfenniger1998,
       author = {{Oll\'e}, Merce and {Pfenniger}, Daniel},
        title = "{Vertical orbital structure around the Lagrangian points in barred galaxies. Link with the secular evolution of galaxies}",
      journal = {\aap},
         year = 1998,
        month = jun,
       volume = {334},
        pages = {829-839},
       adsurl = {https://ui.adsabs.harvard.edu/abs/1998A&A...334..829O}
}

@ARTICLE{Patsis2006,
       author = {{Patsis}, P.~A.},
        title = "{The stellar dynamics of spiral arms in barred spiral galaxies}",
      journal = {\mnras},
         year = 2006,
        month = jun,
       volume = {369},
       number = {1},
        pages = {L56-L60},
          doi = {10.1111/j.1745-3933.2006.00174.x},
       adsurl = {https://ui.adsabs.harvard.edu/abs/2006MNRAS.369L..56P}
}

@BOOK{Patsis2009,
  author    = {Patsis, P. A.},
  title     = {Chaos in Astronomy},
  publisher = {Springer},
  year      = {2009},
  editor    = {Contopoulos, G. and Patsis, P. A.},
  series    = {Astrophysics and Space Science Proceedings},
  volume    = {8},
  doi       = {10.1007/978-3-540-75826-6}
}

@ARTICLE{Patsis2010,
       author = {{Patsis}, P.~A. and {Kalapotharakos}, C. and {Grosb{\o}l}, P.},
        title = "{NGC1300 dynamics - III. Orbital analysis}",
      journal = {\mnras},
         year = 2010,
        month = oct,
       volume = {408},
       number = {1},
        pages = {22-39},
          doi = {10.1111/j.1365-2966.2010.17062.x},
archivePrefix = {arXiv},
       eprint = {1009.0403},
 primaryClass = {astro-ph.CO},
       adsurl = {https://ui.adsabs.harvard.edu/abs/2010MNRAS.408...22P}
}

@ARTICLE{PatsisTsigaridi2017,
       author = {{Patsis}, P.~A. and {Tsigaridi}, L.},
        title = "{The flow in the spiral arms of slowly rotating bar-spiral models}",
      journal = {\apss},
         year = 2017,
        month = jul,
       volume = {362},
       number = {7},
          eid = {129},
        pages = {129},
          doi = {10.1007/s10509-017-3109-9},
       adsurl = {https://ui.adsabs.harvard.edu/abs/2017Ap&SS.362..129P}
}

@ARTICLE{Pfenninger1984b,
  author  = {{Pfenniger}, D.},
  title   = "{The dynamics of barred galaxies}",
  journal = {\aap},
  year    = 1984,
  volume  = {141},
  pages   = {171-183},
  adsurl  = {https://ui.adsabs.harvard.edu/abs/1984A&A...141..171P}
}

@ARTICLE{Rautiainen1999,
       author = {{Rautiainen}, P. and {Salo}, H.},
        title = "{Multiple pattern speeds in barred galaxies. I. Two-dimensional models}",
      journal = {\aap},
         year = 1999,
        month = aug,
       volume = {348},
        pages = {737-754},
       adsurl = {https://ui.adsabs.harvard.edu/abs/1999A&A...348..737R}
}

@ARTICLE{Rautiainen2000,
       author = {{Rautiainen}, P. and {Salo}, H.},
        title = "{N-body simulations of resonance rings in galactic disks}",
      journal = {\aap},
         year = 2000,
        month = oct,
       volume = {362},
        pages = {465-586},
       adsurl = {https://ui.adsabs.harvard.edu/abs/2000A&A...362..465R}
}

@ARTICLE{RocaFabrega2013,
       author = {{Roca-F{\`a}brega}, Santi and {Valenzuela}, Octavio and {Figueras}, Francesca and {Romero-G{\'o}mez}, Merc{\`e} and {Vel{\'a}zquez}, H{\'e}ctor and {Antoja}, Teresa and {Pichardo}, B{\'a}rbara},
        title = "{On galaxy spiral arms' nature as revealed by rotation frequencies}",
      journal = {\mnras},
         year = 2013,
        month = jul,
       volume = {432},
       number = {4},
        pages = {2878-2885},
          doi = {10.1093/mnras/stt643},
archivePrefix = {arXiv},
       eprint = {1302.6981},
 primaryClass = {astro-ph.GA},
       adsurl = {https://ui.adsabs.harvard.edu/abs/2013MNRAS.432.2878R}
}

@ARTICLE{MRG2006,
       author = {{Romero-G{\'o}mez}, M. and {Masdemont}, J.~J. and {Athanassoula}, E. and {Garc{\'\i}a-G{\'o}mez}, C.},
        title = "{The origin of rR$_{1}$ ring structures in barred galaxies}",
      journal = {\aap},
         year = 2006,
        month = jul,
       volume = {453},
       number = {1},
        pages = {39-45},
          doi = {10.1051/0004-6361:20054653},
archivePrefix = {arXiv},
       eprint = {astro-ph/0603124},
 primaryClass = {astro-ph},
       adsurl = {https://ui.adsabs.harvard.edu/abs/2006A&A...453...39R}
}

@ARTICLE{MRG2007,
       author = {{Romero-G{\'o}mez}, M. and {Athanassoula}, E. and {Masdemont}, J.~J. and {Garc{\'\i}a-G{\'o}mez}, C.},
        title = "{The formation of spiral arms and rings in barred galaxies}",
      journal = {\aap},
         year = 2007,
        month = sep,
       volume = {472},
       number = {1},
        pages = {63-75},
          doi = {10.1051/0004-6361:20077504},
archivePrefix = {arXiv},
       eprint = {0705.2958},
 primaryClass = {astro-ph},
       adsurl = {https://ui.adsabs.harvard.edu/abs/2007A&A...472...63R}
}

@ARTICLE{Roskar2012,
       author = {{Ro{\v{s}}kar}, Rok and {Debattista}, Victor P. and {Quinn}, Thomas R. and {Wadsley}, James},
        title = "{Radial migration in disc galaxies - I. Transient spiral structure and dynamics}",
      journal = {\mnras},
         year = 2012,
        month = nov,
       volume = {426},
       number = {3},
        pages = {2089-2106},
          doi = {10.1111/j.1365-2966.2012.21860.x},
archivePrefix = {arXiv},
       eprint = {1110.4413},
 primaryClass = {astro-ph.GA},
       adsurl = {https://ui.adsabs.harvard.edu/abs/2012MNRAS.426.2089R}
}

@ARTICLE{SanchezMartin2023,
       author = {{S{\'a}nchez-Mart{\'\i}n}, P. and {Garc{\'\i}a-G{\'o}mez}, C. and {Masdemont}, J.~J. and {Romero-G{\'o}mez}, M.},
        title = "{Formation of asymmetric arms in barred galaxies}",
      journal = {\mnras},
         year = 2023,
        month = apr,
       volume = {520},
       number = {3},
        pages = {3909-3915},
          doi = {10.1093/mnras/stad303},
archivePrefix = {arXiv},
       eprint = {2301.11385},
 primaryClass = {astro-ph.GA},
       adsurl = {https://ui.adsabs.harvard.edu/abs/2023MNRAS.520.3909S}
}

@ARTICLE{SanchezMartin2026,
       author = {{S{\'a}nchez-Mart{\'\i}n}, P. and {L{\'o}pez-Vilamaj{\'o}}, M. and {Romero-G{\'o}mez}, M. and {Masdemont}, J.~J.},
        title = "{Arm morphology in off-centre barred galaxies}",
      journal = {arXiv e-prints},
         year = 2026,
        month = may,
          eid = {arXiv:2605.19882},
        pages = {arXiv:2605.19882},
archivePrefix = {arXiv},
       eprint = {2605.19882},
 primaryClass = {astro-ph.GA},
          doi = {10.48550/arXiv.2605.19882},
       adsurl = {https://ui.adsabs.harvard.edu/abs/2026arXiv260519882S}
}

@ARTICLE{SellwoodCarlberg1984,
       author = {{Sellwood}, J.~A. and {Carlberg}, R.~G.},
        title = "{Spiral instabilities provoked by accretion and star formation}",
      journal = {\apj},
         year = 1984,
        month = jul,
       volume = {282},
        pages = {61-74},
          doi = {10.1086/162176},
       adsurl = {https://ui.adsabs.harvard.edu/abs/1984ApJ...282...61S}
}

@ARTICLE{Sellwood&Wilkinson1993,
       author = {{Sellwood}, J.~A. and {Wilkinson}, A.},
        title = "{Dynamics of barred galaxies}",
      journal = {Reports on Progress in Physics},
         year = 1993,
        month = feb,
       volume = {56},
       number = {2},
        pages = {173-256},
          doi = {10.1088/0034-4885/56/2/001},
archivePrefix = {arXiv},
       eprint = {astro-ph/0608665},
 primaryClass = {astro-ph},
       adsurl = {https://ui.adsabs.harvard.edu/abs/1993RPPh...56..173S}
}

@ARTICLE{Sellwood2011,
       author = {{Sellwood}, J.~A.},
        title = "{The lifetimes of spiral patterns in disc galaxies}",
      journal = {\mnras},
         year = 2011,
        month = jan,
       volume = {410},
       number = {3},
        pages = {1637-1646},
          doi = {10.1111/j.1365-2966.2010.17545.x},
archivePrefix = {arXiv},
       eprint = {1008.2737},
 primaryClass = {astro-ph.CO},
       adsurl = {https://ui.adsabs.harvard.edu/abs/2011MNRAS.410.1637S}
}

@ARTICLE{SellwoodCarlberg2014,
       author = {{Sellwood}, J.~A. and {Carlberg}, R.~G.},
        title = "{Transient Spirals as Superposed Instabilities}",
      journal = {\apj},
         year = 2014,
        month = apr,
       volume = {785},
       number = {2},
          eid = {137},
        pages = {137},
          doi = {10.1088/0004-637X/785/2/137},
archivePrefix = {arXiv},
       eprint = {1403.1135},
 primaryClass = {astro-ph.GA},
       adsurl = {https://ui.adsabs.harvard.edu/abs/2014ApJ...785..137S}
}

@ARTICLE{SellwoodCarlberg2019,
       author = {{Sellwood}, J.~A. and {Carlberg}, Ray G.},
        title = "{Spiral instabilities: mechanism for recurrence}",
      journal = {\mnras},
         year = 2019,
        month = oct,
       volume = {489},
       number = {1},
        pages = {116-131},
          doi = {10.1093/mnras/stz2132},
archivePrefix = {arXiv},
       eprint = {1906.04191},
 primaryClass = {astro-ph.GA},
       adsurl = {https://ui.adsabs.harvard.edu/abs/2019MNRAS.489..116S}
}

@ARTICLE{SellwoodMasters2022,
       author = {{Sellwood}, J.~A. and {Masters}, Karen L.},
        title = "{Spirals in Galaxies}",
      journal = {\araa},
         year = 2022,
        month = aug,
       volume = {60},
          doi = {10.1146/annurev-astro-052920-104505},
archivePrefix = {arXiv},
       eprint = {2110.05615},
 primaryClass = {astro-ph.GA},
       adsurl = {https://ui.adsabs.harvard.edu/abs/2022ARA&A..60...73S}
}

@ARTICLE{Sheth2002,
       author = {{Sheth}, Kartik and {Vogel}, Stuart N. and {Regan}, Michael W. and {Teuben}, Peter J. and {Harris}, Andrew I. and {Thornley}, Michele D.},
        title = "{Molecular Gas and Star Formation in Bars of Nearby Spiral Galaxies}",
      journal = {\aj},
         year = 2002,
        month = nov,
       volume = {124},
       number = {5},
        pages = {2581-2599},
          doi = {10.1086/343835},
archivePrefix = {arXiv},
       eprint = {astro-ph/0208018},
 primaryClass = {astro-ph},
       adsurl = {https://ui.adsabs.harvard.edu/abs/2002AJ....124.2581S}
}

@ARTICLE{Skokos2002a,
       author = {{Skokos}, Ch. and {Patsis}, P.~A. and {Athanassoula}, E.},
        title = "{Orbital dynamics of three-dimensional bars - I. The backbone of three-dimensional bars. A fiducial case}",
      journal = {\mnras},
         year = 2002,
        month = jul,
       volume = {333},
       number = {4},
        pages = {847-860},
          doi = {10.1046/j.1365-8711.2002.05468.x},
archivePrefix = {arXiv},
       eprint = {astro-ph/0204077},
 primaryClass = {astro-ph},
       adsurl = {https://ui.adsabs.harvard.edu/abs/2002MNRAS.333..847S}
}

@ARTICLE{Skokos2002b,
       author = {{Skokos}, Ch. and {Patsis}, P.~A. and {Athanassoula}, E.},
        title = "{Orbital dynamics of three-dimensional bars - II. Investigation of the parameter space}",
      journal = {\mnras},
         year = 2002,
        month = jul,
       volume = {333},
       number = {4},
        pages = {861-870},
          doi = {10.1046/j.1365-8711.2002.05469.x},
archivePrefix = {arXiv},
       eprint = {astro-ph/0204078},
 primaryClass = {astro-ph},
       adsurl = {https://ui.adsabs.harvard.edu/abs/2002MNRAS.333..861S}
}

@ARTICLE{TST2026a,
  author  = {Soler-Terricabras, T. and Romero-G{\'o}mez, M. and Roca-F{\`a}brega, S.},
  title   = "{Invariant manifolds in barred galaxy simulations - I. Material density waves}",
  journal = {\aap},
  year    = {2026},
  volume   = {710},
  pages    = {A128},
  eid      = {A128},
  doi      = {10.1051/0004-6361/202659896}
}

@ARTICLE{TST2026b,
  author  = {Soler-Terricabras, T. and Romero-G{\'o}mez, M. and Roca-F{\`a}brega, S.},
  title   = "{Invariant manifolds in barred galaxy simulations - II. Quantitative evidence of manifold-trapping in spiral arm formation}",
  journal = {\aap},
  year    = {2026},
  volume   = {711},
  pages    = {A180},
  eid      = {A180},
  doi      = {10.1051/0004-6361/202660208}
}

@ARTICLE{TsigaridiPatsis2013,
       author = {{Tsigaridi}, L. and {Patsis}, P.~A.},
        title = "{The backbones of stellar structures in barred-spiral models - the concerted action of various dynamical mechanisms on galactic discs}",
      journal = {\mnras},
         year = 2013,
        month = oct,
       volume = {434},
       number = {4},
        pages = {2922-2939},
          doi = {10.1093/mnras/stt1207},
       adsurl = {https://ui.adsabs.harvard.edu/abs/2013MNRAS.434.2922T}
}

@ARTICLE{TsigaridiPatsis2015,
       author = {{Tsigaridi}, L. and {Patsis}, P.~A.},
        title = "{Morphologies introduced by bistability in barred-spiral galactic potentials}",
      journal = {\mnras},
         year = 2015,
        month = apr,
       volume = {448},
       number = {4},
        pages = {3081-3092},
          doi = {10.1093/mnras/stv206},
archivePrefix = {arXiv},
       eprint = {1502.00548},
 primaryClass = {astro-ph.GA},
       adsurl = {https://ui.adsabs.harvard.edu/abs/2015MNRAS.448.3081T}
}

@ARTICLE{Tsoutsis2008,
       author = {{Tsoutsis}, P. and {Efthymiopoulos}, C. and {Voglis}, N.},
        title = "{The coalescence of invariant manifolds and the spiral structure of barred galaxies}",
      journal = {\mnras},
         year = 2008,
        month = jul,
       volume = {387},
       number = {3},
        pages = {1264-1280},
          doi = {10.1111/j.1365-2966.2008.13331.x},
archivePrefix = {arXiv},
       eprint = {0804.2376},
 primaryClass = {astro-ph},
       adsurl = {https://ui.adsabs.harvard.edu/abs/2008MNRAS.387.1264T}
}

@ARTICLE{Tsoutsis2009,
       author = {{Tsoutsis}, P. and {Kalapotharakos}, C. and {Efthymiopoulos}, C. and {Contopoulos}, G.},
        title = "{Invariant manifolds and the response of spiral arms in barred galaxies}",
      journal = {\aap},
         year = 2009,
        month = mar,
       volume = {495},
       number = {3},
        pages = {743-758},
          doi = {10.1051/0004-6361:200810149},
archivePrefix = {arXiv},
       eprint = {0811.4521},
 primaryClass = {astro-ph},
       adsurl = {https://ui.adsabs.harvard.edu/abs/2009A&A...495..743T}
}

@ARTICLE{Vasiliev2019,
       author = {{Vasiliev}, Eugene},
        title = "{AGAMA: action-based galaxy modelling architecture}",
      journal = {\mnras},
         year = 2019,
        month = jan,
       volume = {482},
       number = {2},
        pages = {1525-1544},
          doi = {10.1093/mnras/sty2672},
archivePrefix = {arXiv},
       eprint = {1802.08239},
 primaryClass = {astro-ph.GA},
       adsurl = {https://ui.adsabs.harvard.edu/abs/2019MNRAS.482.1525V}
}

@ARTICLE{Voglis2006a,
       author = {{Voglis}, N. and {Stavropoulos}, I. and {Kalapotharakos}, C.},
        title = "{Chaotic motion and spiral structure in self-consistent models of rotating galaxies}",
      journal = {\mnras},
         year = 2006,
        month = oct,
       volume = {372},
       number = {2},
        pages = {901-922},
          doi = {10.1111/j.1365-2966.2006.10914.x},
archivePrefix = {arXiv},
       eprint = {astro-ph/0606561},
 primaryClass = {astro-ph},
       adsurl = {https://ui.adsabs.harvard.edu/abs/2006MNRAS.372..901V}
}

@ARTICLE{Wada2011,
       author = {{Wada}, Keiichi and {Baba}, Junichi and {Saitoh}, Takayuki R.},
        title = "{Interplay between Stellar Spirals and the Interstellar Medium in Galactic Disks}",
      journal = {\apj},
         year = 2011,
        month = jul,
       volume = {735},
       number = {1},
          eid = {1},
        pages = {1},
          doi = {10.1088/0004-637X/735/1/1},
archivePrefix = {arXiv},
       eprint = {1104.1287},
 primaryClass = {astro-ph.GA},
       adsurl = {https://ui.adsabs.harvard.edu/abs/2011ApJ...735....1W}
}

@ARTICLE{Williams2024,
       author = {{Williams}, Thomas G. and {Lee}, Janice C. and {Larson}, Kirsten L. and {Leroy}, Adam K. and {Sandstrom}, Karin and {Schinnerer}, Eva and {Thilker}, David A. and {Belfiore}, Francesco and {Egorov}, Oleg V. and {Rosolowsky}, Erik and {Sutter}, Jessica and {DePasquale}, Joseph and {Pagan}, Alyssa and {Berger}, Travis A. and {Anand}, Gagandeep S. and {Barnes}, Ashley T. and {Bigiel}, Frank and {Boquien}, M{\'e}d{\'e}ric and {Cao}, Yixian and {Chastenet}, J{\'e}r{\'e}my and {Chevance}, M{\'e}lanie and {Chown}, Ryan and {Dale}, Daniel A. and {Deger}, Sinan and {Eibensteiner}, Cosima and {Emsellem}, Eric and {Faesi}, Christopher M. and {Glover}, Simon C.~O. and {Grasha}, Kathryn and {Hannon}, Stephen and {Hassani}, Hamid and {Henshaw}, Jonathan D. and {Jim{\'e}nez-Donaire}, Mar{\'\i}a J. and {Kim}, Jaeyeon and {Klessen}, Ralf S. and {Koch}, Eric W. and {Li}, Jing and {Liu}, Daizhong and {Meidt}, Sharon E. and {M{\'e}ndez-Delgado}, J. Eduardo and {Murphy}, Eric J. and {Neumann}, Justus and {Neumann}, Lukas and {Neumayer}, Nadine and {Oakes}, Elias K. and {Pathak}, Debosmita and {Pety}, J{\'e}r{\^o}me and {Pinna}, Francesca and {Querejeta}, Miguel and {Ramambason}, Lise and {Romanelli}, Andrea and {Sormani}, Mattia C. and {Stuber}, Sophia K. and {Sun}, Jiayi and {Teng}, Yu-Hsuan and {Usero}, Antonio and {Watkins}, Elizabeth J. and {Weinbeck}, Tony D.},
        title = "{PHANGS-JWST: Data-processing Pipeline and First Full Public Data Release}",
      journal = {\apjs},
         year = 2024,
        month = jul,
       volume = {273},
       number = {1},
          eid = {13},
        pages = {13},
          doi = {10.3847/1538-4365/ad4be5},
archivePrefix = {arXiv},
       eprint = {2401.15142},
 primaryClass = {astro-ph.GA},
       adsurl = {https://ui.adsabs.harvard.edu/abs/2024ApJS..273...13W}
}

@ARTICLE{Zouloumi2024,
       author = {{Zouloumi}, K. and {Harsoula}, M. and {Efthymiopoulos}, C.},
        title = "{Multiple pattern speeds and the manifold spirals in a simulation of a barred spiral galaxy}",
      journal = {\mnras},
         year = 2024,
        month = apr,
       volume = {529},
       number = {3},
        pages = {1941-1957},
          doi = {10.1093/mnras/stae353},
       adsurl = {https://ui.adsabs.harvard.edu/abs/2024MNRAS.529.1941Z}
}

\begin{appendix}

\section{Manifold-trapping in the unstable interior branches}\label{sec:appendix} 

\begin{figure}[!h]
       \centering
       \includegraphics[width=0.98\hsize]{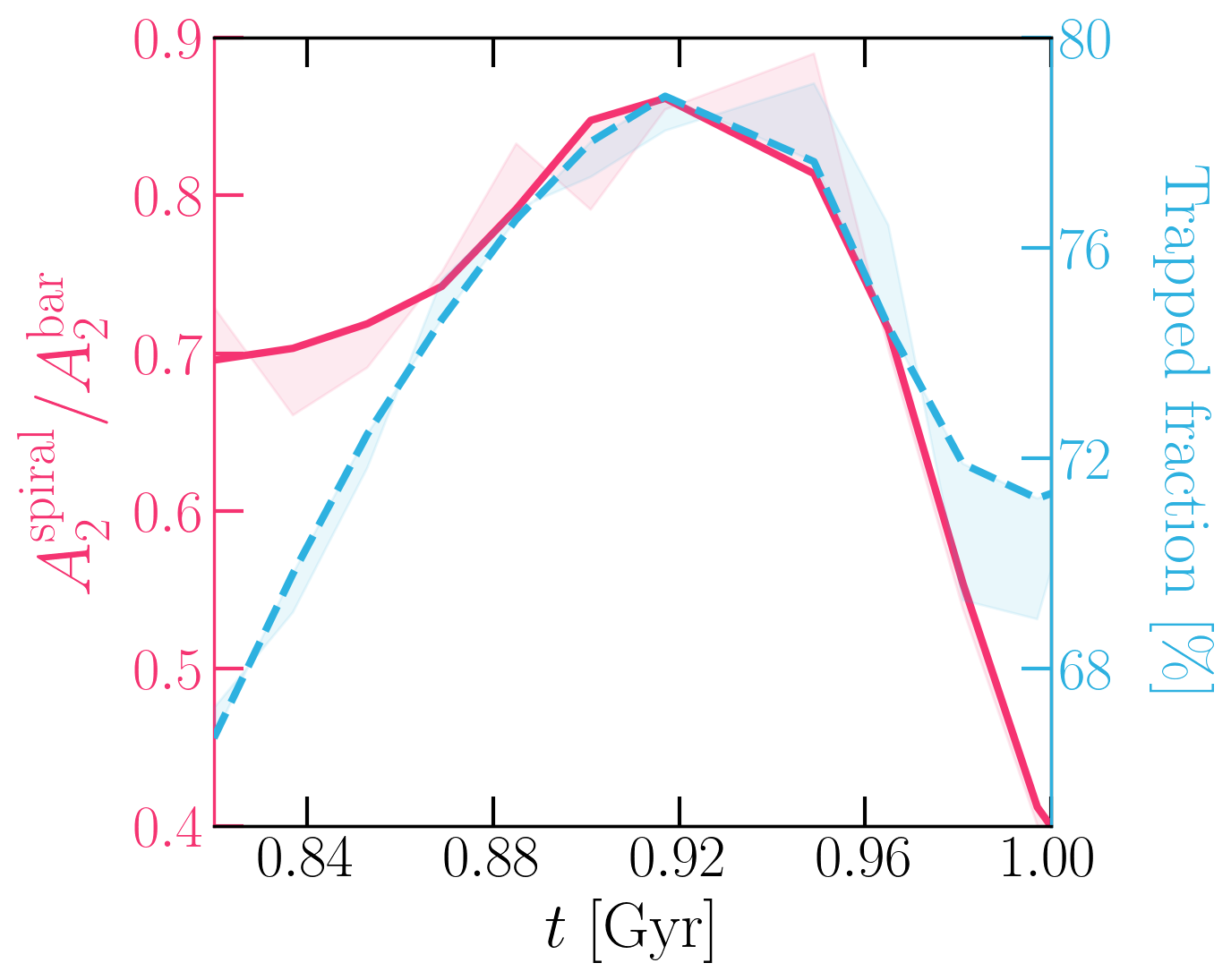}
        \caption{Time evolution of the fraction of trapped particles in the interior unstable branches of the invariant manifolds (blue dashed curve, right axis) over the interval $t = 0.804$ Gyr and $t = 1.000$ Gyr after the initial conditions (strong spiral arms regime).
        This is shown alongside the ratio between the spiral arm strength, quantified via the Fourier amplitude
        $A_2^\text{spiral}$ and the bar strength $A_2^\text{bar}$ (solid magenta curve, left axis).
        Shaded regions correspond to deviations from the smoothed trends.
        }
        \label{fig:trapped}
    \end{figure}

   To examine the behaviour described in Sec.~\ref{sec:inward_flows} in greater detail, Fig.~\ref{fig:trapped} quantifies the effect of the interior unstable branches of the invariant manifolds, $\mathcal{W}_{\ell_i}^{u,i}$, $i \in \{1,2\}$ (see Fig.~\ref{fig:esquema}a), at each simulation snapshot in the regime of strongly developed spiral arms. The quantification follows the same procedure described in Section 2 of \citetalias{TST2026b} —namely, the use of constant-azimuth sections of the manifold branches and the application of the trapping criterion defined therein—with the important distinction that, whereas \citetalias{TST2026b} focused on the exterior unstable branches, here we apply the method to the interior ones only.
   
   The temporal evolution of the trapped fraction in the interior unstable branches is shown in Fig.~\ref{fig:trapped} (light-blue dashed curve, representing the fraction relative to the manifold-compatible population\footnote{As defined in \citetalias{TST2026a} of this series, manifold-compatible particles are those whose Jacobi energy is $E_{L_{1,2}}\leq E_J\leq E_\text{man}$, where $E_{L_{1,2}}$ is that of the saddle equilibrium points $L_1-L_2$, and $E_\text{man}$ is that of the invariant manifold, $E_\text{man}$. This population represents $\sim 30-40\%$ of the disc population.}), together with the ratio between the spiral arm strength, traced by the $m = 2$ Fourier amplitude $A_2^\text{spiral}$, and the bar strength, $A_2^\text{bar}$ (pink solid curve). 
   As discussed in Sec.~\ref{sec:distortion}, the spiral amplitude $A_2^\mathrm{spiral}$ is averaged over the radial range $[R_1,10]\,\mathrm{kpc}$, whereas the bar amplitude $A_2^\mathrm{bar}$ is averaged over $[R_0,R_1]\,\mathrm{kpc}$, where $R_0$ and $R_1$ respectively denote the inner and outer edges of the bar region as defined in Appendix~B of \citet{Dehnen2022}.
   Both amplitudes therefore probe different regions of the disc.
   A strong correlation is found between these two quantities, with the normalised cross-correlation function reaching $\text{CCF}_\text{max} \approx 0.9$ and no measurable lag between the signals. 
   In contrast to the delayed response discussed for the exterior unstable branches in \citetalias{TST2026b}, no delay is expected here because the bar is already a fully developed self-gravitating structure. 
   In this case, the trapping efficiency does not depend on building up a collective gravitational response from previously untrapped disc particles, but rather on the instantaneous non-axisymmetric potential configuration generated by the relative bar--spiral strength.
   The ratio $A_2^\text{spiral}$/$A_2^\text{bar}$ used here (Fig.~\ref{fig:trapped}, magenta solid curve) primarily measures the instantaneous relative dominance of the spiral perturbation with respect to the bar. 
   
   We find that the trapping efficiency of the interior unstable branches increases precisely during phases in which the spiral perturbation becomes more dominant relative to the bar, indicating that the geometry and dynamical activity of these branches are tightly coupled to the instantaneous non-axisymmetric potential configuration.
   This behaviour provides clear evidence that the mechanism proposed in Fig.~\ref{fig:esquema} is indeed at work: the self-gravity of the spiral arms modifies the manifold structure so that the interior unstable branches become dynamically active overlapping with the inner regions of the arms, and their increasing trapping efficiency contributes to the subsequent weakening of the spiral arms through inflows of particles towards the bar region.
   
   This interpretation is further supported by a closer inspection of the strong-spiral-arm regime ($0.8\,\mathrm{Gyr}\lesssim t \lesssim 1.0\,\mathrm{Gyr}$), comparing simultaneously Fig.~\ref{fig:psi} and Fig.~\ref{fig:trapped}. 
   Three different phases can be identified during this interval.
   At early times ($0.8\,\mathrm{Gyr}\lesssim t \lesssim 0.9\,\mathrm{Gyr}$), both the bar and the spiral arms exhibit large amplitudes ($A_2^\mathrm{bar}\sim0.5$, $A_2^\mathrm{spiral}\sim0.3$), coinciding with the largest distortions of the equilibrium points. 
   Around $t\approx 0.9\,\mathrm{Gyr}$, the relative contribution of the spiral perturbation becomes maximal ($A_2^\mathrm{bar}\sim0.3$, $A_2^\mathrm{spiral}\sim0.25$), and this is precisely when the trapping efficiency within the interior unstable branches reaches its peak. 
   Subsequently ($0.9\,\mathrm{Gyr}\lesssim t \lesssim 1.0\,\mathrm{Gyr}$), the bar amplitude remains approximately constant and the spiral arms weaken significantly ($A_2^\mathrm{spiral}\sim0.1$); at the same time, both the trapping efficiency and the equilibrium-point distortions decrease accordingly.
   Overall, these results indicate that the inflow efficiency associated with the interior unstable branches is controlled not simply by the absolute spiral amplitude, but by the relative dynamical dominance of the spiral perturbation with respect to the bar within the total non-axisymmetric potential.
   In this sense, the ratio $A_2^\mathrm{spiral}/A_2^\mathrm{bar}$ provides a direct measure of the instantaneous bar--spiral competition that governs the activation of the interior unstable branches in the spiral-arm weakening mechanism.

\section{Inflows in the standard manifold configuration}\label{sec:no_inflow} 
    
    \begin{figure*}[!ht]
       \centering
       \includegraphics[width=\hsize]{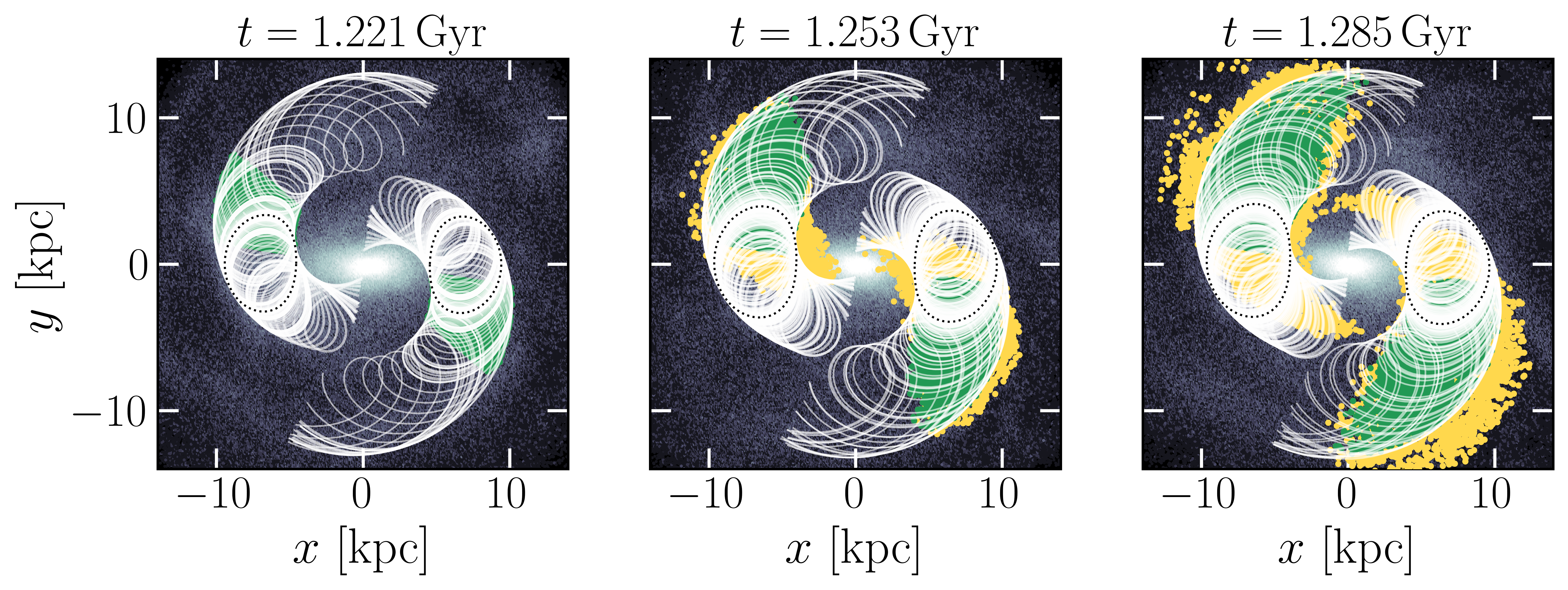}
        \caption{
        Time evolution of the bundle of particles initially trapped within the exterior unstable branches at $t=1.221$~Gyr, within the angular interval $10^\circ \leq \theta \leq 40^\circ$. A total of 12,117 particles are overlaid in each panel (4,728 associated with the $L_1$ exterior unstable branch and 7,389 with the $L_2$ branch). Subsequent panels follow the same particles forward in time, illustrating the evolution of their trajectories and trapping state, as defined in \citetalias{TST2026b}: green dots correspond to particles that remain trapped on the exterior branches (IN--IN) relative to the initial selection, while yellow dots indicate particles that escape from the trapped state (IN--OUT). White curves show the unstable branches of the invariant manifolds, including both the interior and exterior branches; the interior branches extend toward the bar region, whereas the exterior branches spread out into the spiral arms. The associated Lyapunov orbits are shown as black dotted curves. The background displays the face-on surface density of the disc, and each panel is labelled with the corresponding simulation time. {In this regime, the equilibrium points $L_1$ and $L_2$ are aligned with the bar semi-major axis (standard configuration; see Fig.~\ref{fig:esquema}b). The outward flows contribute to the spiral arm structure, while the inward flows contribute to the inner ring.}}
        \label{fig:no_inward_flow}
    \end{figure*}

    {Figure~\ref{fig:no_inward_flow} shows the evolution of a bundle of particles selected as in Fig.~\ref{fig:inward_flow} but now in the standard configuration (during the time interval 1.221~Gyr $\leq t \leq$ 1.285~Gyr). 
    As discussed in \citetalias{TST2026a} and \citetalias{TST2026b} of this series, the outward propagation of particles along the exterior unstable branches contributes to the spiral arm structure. 
    Although it is indeed the case in Fig.~\ref{fig:no_inward_flow}, some particles sufficiently close to the chaotic regions around $L_1$ and $L_2$ move inwards instead, directly contributing to the formation of the inner ring (see $t=1.285$~Gyr, Fig.~\ref{fig:no_inward_flow}). 
    Note that, in this standard configuration, these inflows do not directly populate the bar itself, but rather the ring surrounding it, as expected from \citet{MRG2006, MRG2007}.
     The limited contribution of these inflows to the inner ring is a direct consequence of the alignment between the equilibrium points and the bar: as long as $L_1$ and $L_2$ remain close to the bar ends, they are spatially separated from the density peak of the spiral arms, so that only a small fraction of the manifold-compatible population lies sufficiently close to the chaotic regions around $L_1$ and $L_2$ to be redirected inwards.}

\section{Flux-tube and apocentric manifold formulations}\label{sec:grecs}

{As an additional remark, it is worth placing the results of this work in the context of the two complementary formulations of manifold theory developed in the literature. 
The present work, as well as \citetalias{TST2026a} and \citetalias{TST2026b} of this series, adopts the original flux-tube formulation of invariant manifolds \citep{Koon2000, MRG2006, MRG2007, Athanassoula2009b, Athanassoula2009a, Athanassoula2010, Athanassoula2012}. 
In this picture, spiral arms are identified with the continuous orbital flow guided by the unstable manifolds emerging from the vicinity of the unstable Lagrangian points. 
Within the flux-tube formulation, and as also discussed in \citetalias{TST2026b}, spiral arms can remain prominent provided that the invariant manifolds are continuously fed by material from the bar region through the vicinity of the Lagrangian points \citep{Athanassoula2010}. 
In this view, the longevity of the spiral pattern is conditional on a sustained external supply feeding the manifold tubes, without which the outward, essentially one-way transport described in Section~\ref{sec:weakening} would be expected to gradually deplete the reservoir of manifold-compatible particles.
Consequently, the mechanism identified here does not contradict the original picture proposed by \citet{MRG2006, MRG2007, Athanassoula2009b, Athanassoula2009a, Athanassoula2010}, but rather provides a dynamical origin for the resupply process on which the long-term maintenance of manifold-driven spiral arms depends.}

{An alternative, but dynamically equivalent, formulation is the apocentric-manifold approach \citep{Voglis2006a, Tsoutsis2008, Tsoutsis2009, Harsoula2016}, in which the spiral pattern is assumed to be traced by the locus of the successive apsidal (typically apocentric) passages of manifold-guided chaotic orbits. 
By focusing on the recurrent apocentric positions rather than on the continuous orbital flow, this formulation naturally emphasises the recurrent character of stellar motions along the spiral pattern.}
{As discussed in \citet{Efthymiopoulos2010}, one of the conceptual issues raised against the flux-tube formulation was in the context of spiral arm formation.
In the flux-tube models of rings, homoclinic and heteroclinic connections naturally allow material to return to the inner galaxy through the exterior stable manifold branches \citep[see][]{MRG2006, MRG2007}. 
By contrast, spiral arms lack a comparable overdensity at their outer end, so that the standard flux-tube picture appears to imply a predominantly outward transport of material, motivating the need for a replenishment mechanism if the spiral structure is to be long-lived.
The distorted manifold configuration identified in the present work offers a natural resolution of this issue. 
The inward flows developing along the inner parts of the spiral arms return a significant fraction of the manifold-guided material to the bar region, where it may contribute to the chaotic envelope surrounding the bar before becoming available for subsequent spiral episodes.
In this sense, the recurrent character of spiral activity emerges naturally within the flux-tube framework itself, while remaining consistent with the recurrent orbital behaviour emphasised by the apocentric-manifold formulation.}

{More generally, the present series of papers suggests that manifold dynamics should not be regarded as operating in isolation from the classical density-wave response of the disc. As quantified in \citetalias{TST2026a}, the bulk of the stellar disc is composed of low-(Jacobi) energy, nearly circular orbits that are insensitive to the manifold dynamics in a direct way. 
In contrast, the manifold-compatible population occupies a higher-energy phase-space component. 
\citetalias{TST2026b} demonstrated that these two orbital populations are dynamically coupled, with the manifold-guided material perturbing the low-energy disc population, which in turn provides the dominant density response outlining the spiral pattern. This led us to propose the novel concept of \textit{material density waves}, in which the observed spiral arms arise from the combined action of these two dynamically distinct, yet mutually interacting, kinematic populations. 
Within this interpretation, invariant manifolds remain the driver of the spiral perturbation, while the visible spiral pattern is largely built by the response of the low-energy disc. 
We therefore view this hybrid picture as a natural extension of the flux-tube manifold theory, capable of reconciling the manifold framework with the well-established orbital populations present in self-consistent barred galaxies. 
In particular, while the apocentric-manifold formulation may capture the recurrent behaviour of manifold-guided orbits, it does not explicitly address how these orbits interact with the dynamically colder, low-energy population that dominates the stellar disc. 
Our results suggest that this coupling is a key ingredient for understanding the formation and persistence of spiral structure in self-consistent barred galaxies.}

\end{appendix}

\end{document}